\documentclass[11pt,a4paper]{article}
\usepackage{authblk}
\usepackage[english]{babel}
\usepackage{times}
\usepackage{graphicx}
\usepackage[font=small]{caption}
\usepackage{subfigure}
\usepackage[section]{placeins}
\usepackage{fancyhdr}
\usepackage[dvipsnames]{xcolor}
\usepackage{pst-all}
\usepackage{latexsym}
\usepackage{amsmath}
\usepackage{amssymb}
\usepackage{amsfonts}

\begin{document}

\title{Stress concentration in periodically rough hertzian contact: Hertz to soft-flat-punch transition}
\author[1,2,$^{\dagger}$]{Ren\'{e} Ledesma-Alonso}
\author[1]{Elie Rapha\"{e}l}
\author[2]{Liliane L\'{e}ger}
\author[2]{Frederic Restagno}
\author[2]{Christophe Poulard}
\affil[1]{Laboratoire de Physico-Chimie Th\'{e}orique, UMR CNRS 7083 Gulliver, ESPCI ParisTech, PSL Research University, 10 Rue Vauquelin, 75005 Paris, France}
\affil[2]{Laboratoire de Physique des Solides, CNRS, Univ. Paris-Sud, Universit\'e Paris-Saclay, 91400 Orsay, France}
\affil[$^{\dagger}$]{e-mail: rene.ledesma-alonso@espci.fr}
\date{\today}
\maketitle

\abstract{

We report on the elastic contact between a spherical lens and a patterned substrate, composed of an hexagonal lattice of cylindrical pillars.
The stress field and the size of the contact area are obtained by means of numerical methods: a superposition method of discrete pressure elements and an iterative bisection-like method.
For small indentations, a transition from a Hertzian to a soft-flat-punch behavior is observed when the surface fraction of the substrate that is covered by pillars is increased.
In particular, we present a master curve defined by two dimensionless parameters, which allows to predict the stress at the center of contact region in terms of the surface fraction occupied by pillars.
The transition between the limiting contact regimes, Hertzian and soft-flat-punch, is well described by a rational function.
Additionally, a simple model to describe the Boussines-Cerruti-like contact between the lens and a single elastic pillar, which takes into account the pillar geometry and the elastic properties of the two bodies, is presented.

}


\section{Introduction}
\label{Sec:Intro}

The contact mechanics between an elastic sphere and a flat elastic solid, with frictionless, perfectly smooth and non-conforming surfaces, is beautifully provided by the Hertz theory~\cite{Johnson}.
From the knowledge of the geometry (sphere radius $R$), the elastic properties (Young modulus and Poisson's ratio) and the total load $F$ compressing the two elastic solids, one can forecast the radius of contact $a\sim\sqrt[3]{RF}$, the total indentation $\zeta\sim\sqrt[3]{F^2/R}$ of the two bodies, and the maximum pressure $\sigma_{0H}\sim\sqrt[3]{F/R^2}$.
When adhesion between the two bodies in contact is considered, several modifications from the Hertz theory have been proposed: 1) the Johnson-Kendall-Roberts (JKR) model, which accounts for the attractive surface forces inside the contact region, predicts a larger contact radius than the Hertzian value and applies for soft materials, 2) the Derjaguin-Muller-Toporov (DMT) model, which considers that the attractive forces act only outside the contact region, infers a Hertzian contact profile but with larger deformations (compared to the Hertz theory) for the same load, and suits for hard materials. Finally the Maugis-Dugdale (MD) model describes an unified way in precisely the transition between the two previous models.
 
However, real surfaces are not smooth and the contact mechanics of two real bodies can bitterly deviate from the Hertzian predictions since only partial contact can occur \cite{Johnson,restagno2002adhesion,Poulard2013}.
Attention has been paid to this problem from two different points of departure, the first based on a statistical approach of real surfaces and the second on the deterministic behavior of periodic surfaces.

On one hand, the statistical approach to the contact mechanics of real surfaces bifurcates into three branches.
First, considering the contact between a plane and an isotropically rough surface composed of non-interacting ``spherical asperities'' with a vertical position given by a distribution of heights (exponential and Gaussian)~\cite{Greenwood1966,Greenwood1967,Greenwood1984}, from which estimations of the number of contact spots, the contact area and the total load were obtained in terms of the apparent contact area, the density of asperities, the radius of curvature of the asperities and a function of the heights distribution.
In particular, it has been found that the number of asperities in contact and the real contact area are both proportional to the total load~\cite{Greenwood1966}, and the ratio of the roughness to the indentation has been proposed as a parameter to account for the influence of the roughness on the contact mechanics~\cite{Greenwood1967,Greenwood1984}.
Additionally, useful empirical elasto-plastic models can be found in the recent literature~\cite{Talke2010,Khonsari2014}, which consider nominally rough surfaces and are based on the above-mentioned statistical approaches.
Second, a similar approach to the previous one was developed, but considering a distribution for the shapes of the asperities, \textit{i.e.} an isotropically rough surface composed of non-interacting ``paraboloid-like asperities'' with a Gaussian distribution of the slopes and curvatures~\cite{Bush1975}.
It has been shown that considering a roughness, given by a bandwidth of wavelengths, the real contact area is strictly proportional to the load and, thus, the stress distribution at the average asperity contact region remains unchanged.
Nevertheless, the two above mentioned approaches break down for large loads, at which the separation between the contact surfaces becomes small and the interaction between adjoining asperities cannot be neglected.
Finally, the surface roughness power spectrum, for instance self-affine fractal surfaces and frozen capillary waves, was employed to estimate the ratio between the real and apparent contact areas~\cite{Persson2006}.
This description is particularly accurate for high loads and it is based on the knowledge of the stress distribution, which depends on the average pressure and the magnification (the ratio between size of the apparent contact area and the shortest wavelength roughness).
It also predicts a linear dependence of the real contact area on the total load, for small loads and small real-to-apparent contact area ratios.
Even though the aforementioned statistical methods provide average quantities and global insight of the rough contact phenomenon, the local stress and displacement fields at the contact spots remain inaccessible since the surface topography may not be persistent from one body to another.

On the other hand, in order to gain insight on the local response of the bodies in contact, periodic contact problems have been addressed~\cite{Johnson1984,Block2008}.
For instance the contact between a flat surface and a well-characterized 1D wavy surface, of wavelength $\lambda$ and amplitude $\delta$, has been addressed by superposition of concentrated normal forces at the surface of an infinite half-space (Boussinesq-Cerruti approach).
In this situation, a close-form solution for the stress and displacement fields is obtained and we should encourage the interested reader to review the aforementioned bibliography.
However, we should mention that for a given line load $F$, the local contact occurs in bands of increasing width $2a$ that obey the scaling $F\sim \delta\sin^2\left(\pi a\lambda \right)$ with a maximum stress $\sigma_0\sim\left(\delta/\lambda\right)\sin\left(\pi a/\lambda\right)$.
Following the same ideas, a well-characterized 2D regular and orthogonal wavy surface was analyzed but only asymptotic behaviors have been discerned.
In particular, for light loads, a local Hertzian behavior is expected at the contact between the plane and each summit of the wavy surface, a nearly circular region of contact of radius $a\sim \sqrt[3]{F}$ is observed.
Despite these efforts, the size of the contact region and the stress field provoked by the contact of rough finite-size surfaces remains not predicted by the discussed deterministic method.

Understanding the contact mechanics of rough surfaces, starting with well-controlled patterned surfaces, is fundamental in order to clarify the effect of pattern covering on friction~\cite{Archard1953,Murarash2011,Brormann2013,Kligerman2014,Tsipenyuk2014,Romero2014}, or to identify the influence of texturation on the adhesive properties of surfaces~\cite{Mittal,Kozlova2006,Poulard2011,Tamelier2012}.
This query has received a lot of attention since patterned surfaces have been extensively used as biomimetic surfaces\cite{gorb2007,bhushan2011,hui2004,zhou2013,crosby2005,murphy2009}.
From this point of view, one of the important issues is the contact formation, which fixes the relationship between the measured adhesion and the preload before the detachment \cite{benz2006deformation,greiner2007adhesion}.

In the simplest case of a spherical lens on well-controlled patterned surface made on elastomeric PDMS with circular bumps and cylindrical pillars, a transition from suspended (the lens touches only the top of the bumps/pillars) to intimate contact (the lens gets in contact not only with the bumps/pillars but also with the underlying substrate in between the patterns) has been observed. The contact formation has been shown to depend strongly on the geometry and elastic properties of the textured surface \cite{Poulard2013,Verneuil2007,Hisler2013,Jagota2010}.
In particular, Verneuil and coworkers have shown that the one leading parameter is the surface fraction $\Phi$ occupied by the pillars, since the critical force, which marks the threshold between both contact regimes, scales as $F_c\sim \Phi^3$. 

\begin{figure}
\centering
\includegraphics[width=0.5\textwidth]{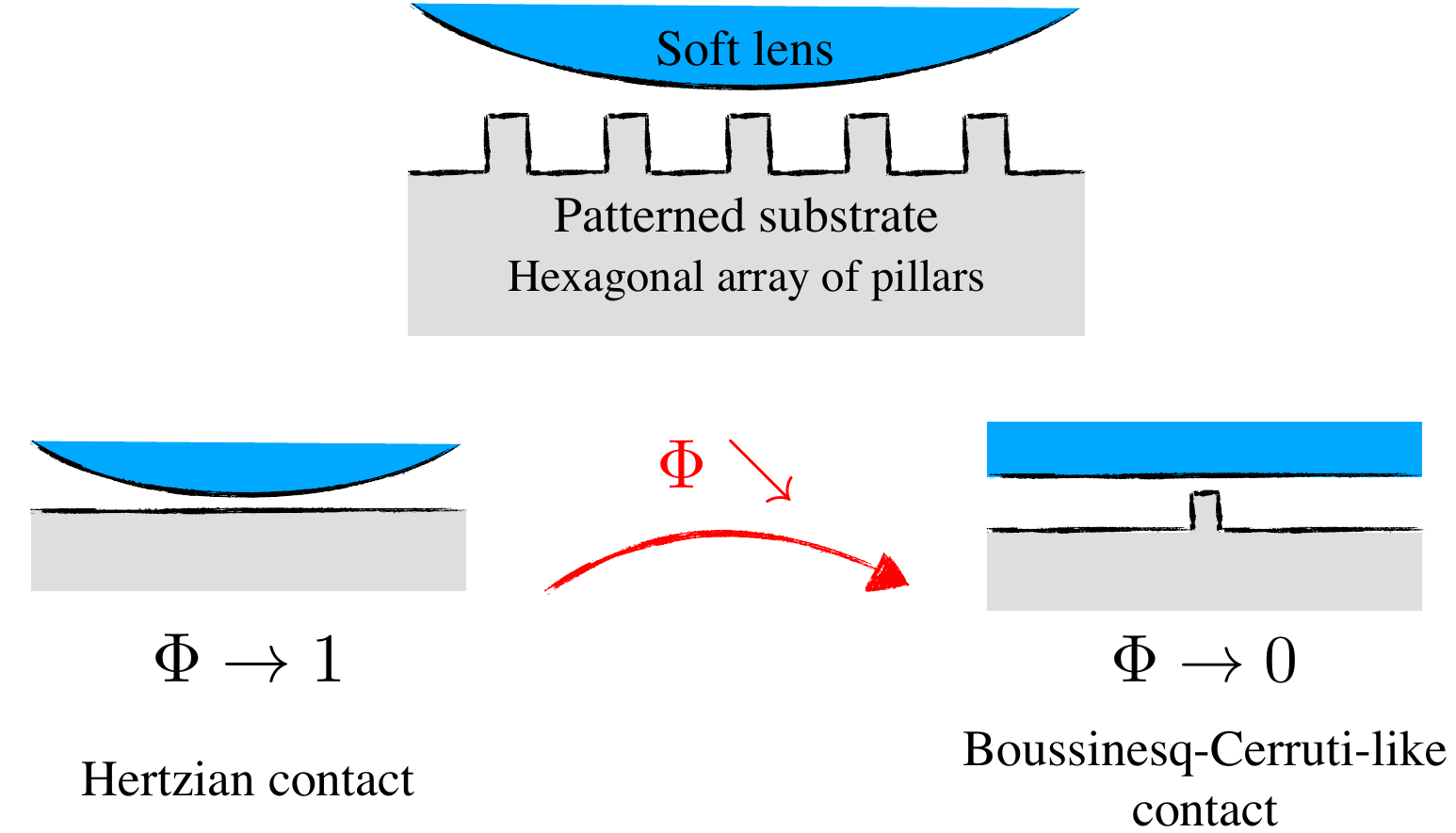}
\caption{Illustration of the discrete contact problem studied in this manuscript and its asymptotic cases, with respect to the surface fraction $\Phi$ of the substrate that is occupied by the pillars.}
\label{Fig:Asymptotic}
\end{figure}

Therefore, there is no doubt about the central role that the surface fraction $\Phi$ plays in the distribution of stresses and in ascertaining the size of the contact region.
The geometry of a sphere in contact with an array of cylindrical pillars is represented graphically in Fig.\ref{Fig:Asymptotic}.
When $\Phi\rightarrow 1$, a classic problem consisting in the contact mechanics of a sphere and a half-space is recovered, for which the stress and displacement fields are described by the Hertz theory.
At the opposite limit, when $\Phi\rightarrow 0$ the contact mechanics is no longer a Hertzian contact but should be represented by the compression of a half-space and a cylindrical pillar, as depicted in Fig.\ref{Fig:Asymptotic}.
The indentation of an elastic half space by a flat rigid punch is described by the classical Boussinesq-Cerruti solution, which is obtained either by the superposition of concentrated normal forces~\cite{Johnson,Love} or by means of a Hankel transform method with an imposed surface displacement~\cite{Sneddon1945,Maugis}.
As well, an analysis, which has remained out of focus, presents a numerical solution that provides the transition from the limit case of the contact between a rigid punch and an elastic half-space to the opposite case of an elastic cylinder and a rigid flat surface~\cite{Gecit1986}.
Nevertheless, the passage from one regime to another is not obvious and work must be done to describe clearly this phenomenon. This is precisely the object of the present article.

In this paper, we discuss the elastic contact between a spherical lens and patterned surface with cylindrical pillars.
The presented analysis, which corresponds to a first approach on the real contact mechanics of periodic rough surfaces, considers two fundamental hypotheses: 1) frictionless contact and 2) no adhesion effects.
The impact of elastic properties and of the geometry of the system is analyzed, with special focus on the surface fraction occupied by the pillars.
A detailed description of the transition from Hertzian contact (short pitch distance) to soft-flat-punch regime (large pitch distance), for the aforementioned lens-patterned substrate system, is presented.
We will discuss the contact behavior for a wide range of elastic parameters (combination of Young's modulus) and displacement of the bodies (total indentation), and analyze the effects of the pitch between pillars in terms of the stress concentration and the number of pillars (real contact area) within the contact region.

\section{System description}
\label{Sec:System}

\begin{figure}[htbp]
\centering
\includegraphics[width=0.6\textwidth]{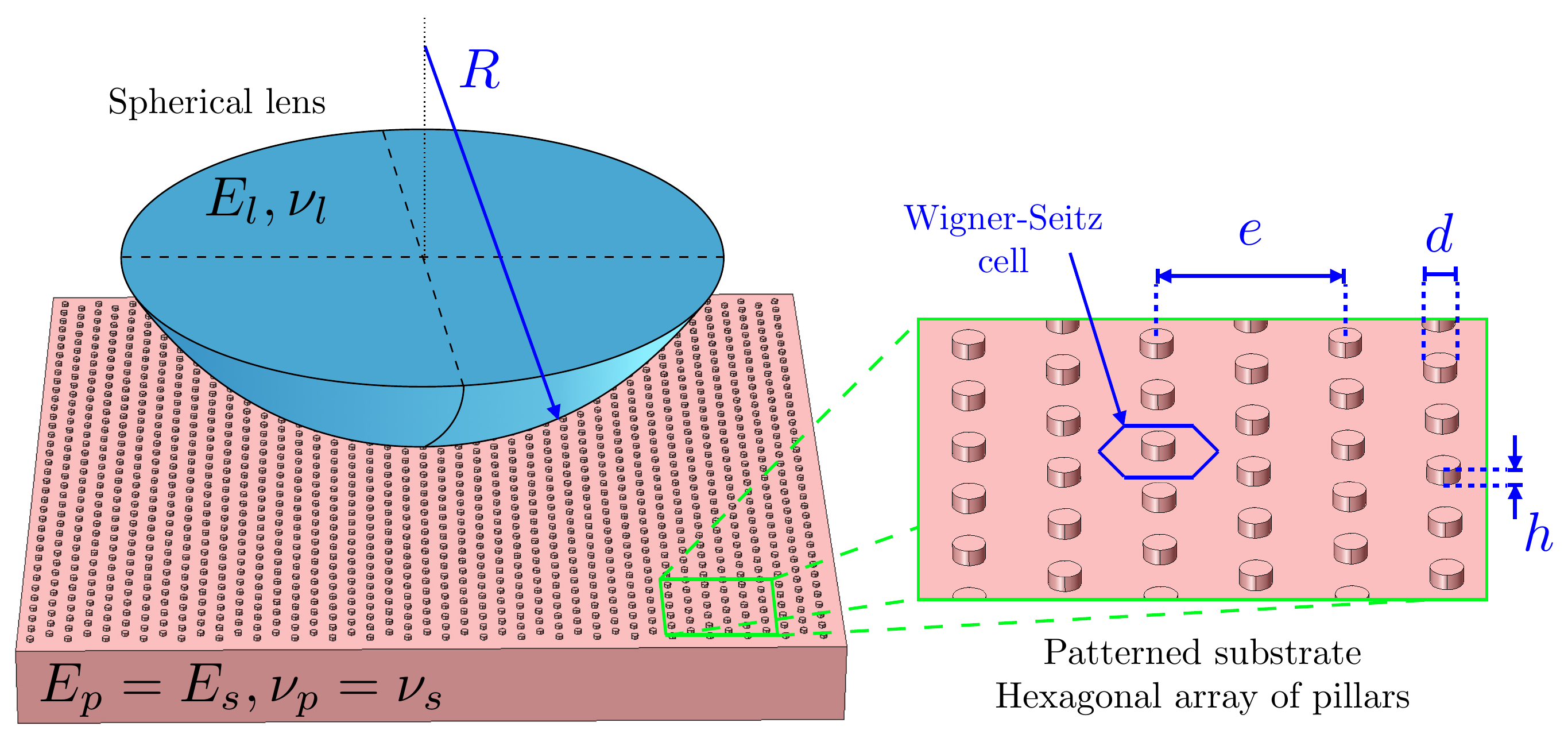}
\caption{Schematic representation of the lens-substrate system. Both bodies are made of elastic materials: the lens is a spherical cap, whereas the substrate consists of a half-space with an hexagonal array of cylindrical pillars at its upper surface. The elastic properties ($E$ Young's modulus and $\nu$ Poisson's ratio) of the components and their relevant dimensions ($R$ lens radius, $d$ diameter and $h$ height of the pillars, and $e$ pitch distance) are shown. The subindices $l$,$p$ and $s$ correspond to lens, pillars and substrate, respectively.}
\label{Fig:Description}
\end{figure}

Consider an elastic system formed by a semi-infinite substrate, which upper surface is patterned with an hexagonal lattice of cylindrical pillars, and a spherical lens.
We imagine a situation, in which the lens is pushed against the substrate, whereas the latter is fixed at the far-field, with respect to the lens initial position.
A discrete mechanical contact region (for small loads) occurs, which produces and conveys stress and displacement fields at the lens and the substrate.
In brief, once a load is applied, the system evolves from an initial undeformed state, at which the lowest point of the lens touches in a single point the center of the top of a central pillar, towards a final deformed state, where multiple contact regions appear.
We expect to observe deviations from this general trend, depending on the specific geometry of the system and the magnitude of the applied load.

Herein, we consider another important hypothesis: the effect of lateral displacement, due to the discrete stress field is negligible with respect to the normal displacement.
Our intuition leads us to imply that the lateral displacement provoked by a single pillar is canceled by the lateral displacement generated by the surrounding pillars and, as a result, the superposition of the effects of all the pillars can be neglected.
In addition, also due to the aforementioned idea, the patterned substrate can be further separated.
To be more precise, the system, represented in Fig.\ref{Fig:Description}, can be decomposed into several independent bodies: 1) a spherical lens, a cap that corresponds to the lower region of a sphere of radius $R$; 2) a flat substrate, which we will take as an elastic half-space; and 3) an array of cylinders, where $d$ and $h$ are the diameter and height of each pillar, whereas $e$ is the pitch, the distance between the center of two pillars.

\subsection{Array of pillars}

When the lens is pushed against the patterned substrate, the number of pillars in contact $Q$ can be estimated by using the Wigner-Seitz cell of an hexagonal lattice of pillars (see Fig.\ref{Fig:Description}). 
Computing the fraction of the surface occupied by the pillars:
\begin{equation}
\Phi=\dfrac{\pi}{2\sqrt{3}}\left(\dfrac{d}{e}\right)^2 \ ,
\label{Eq:Frac}
\end{equation}
and, considering a circular region of radius $a$ as the apparent contact zone, the number of pillars in contact becomes:
\begin{equation}
Q\approx\dfrac{2\pi}{\sqrt{3}}\left(\dfrac{a}{e}\right)^2 \ .
\label{Eq:Count}
\end{equation}
Nevertheless, one cannot predict an accurate number of pillars $Q$ because the apparent contact radius $a$ is unknown \textit{a priori}.
Despite this fact, in the present study we will consider that $R\gg d/2$ and $R\gg a$ which corresponds to a lot of practical situations previously studied in the literature.

Considering a polar system $\left(r,\theta\right)$, the position of the center of the $i$th pillar is given by the coordinates $\left(r_i,\theta_i\right)$, where the value of $i=0$ corresponds to the pillar which center is placed at the origin of the coordinate system.
A detailed description of the pillar array and the precise method to enumerate the pillars and to count the number of pillars in contact is given in the supplementary material~\cite{SM}.
\subsection{Displacement and indentation}

Before the lens and patterned substrate are squeezed and the deformation of both bodies takes place, the center of the spherical lens is placed at $z=h+R$. 
In turn, the equation of the pillars-substrate surface is:
\begin{equation}
g^{\circ}\left(r,\theta\right)=h\sum_{i=0}^{Q}\left[1-H\left(\rho_i-\dfrac{d}{2}\right)\right] \ ,
\end{equation}
where $H$ is a Heaviside step function, taking either the value of 0 if $\rho_i\leq d/2$ or 1 if $\rho_i>d/2$.
As described in the supplementary material~\cite{SM}, relative to the center of the $i$th pillar, which is placed at $\left(r_i,\theta_i\right)$, $\rho_i$ is the distance to the point $\left(r,\theta\right)$, given by:
\begin{equation}
\rho_i=\sqrt{r^2+r_i^2-2r r_i \cos\left(\theta-\theta_i\right)} \ .
\label{Eq:Distnmlk}
\end{equation}
These initial positions indicate that the lens and the patterned substrate are in contact at a single point, at $r=0$ where the lowest position of the lens touches the center of the pillar upper surface.  

Once the load is applied and the system has attained equilibrium, the final gap between the surfaces, the space separating the lens and the pillars-substrate body, is now given by:
\begin{equation}
\Delta\left(r,\theta\right)=h+R-\sqrt{R^2-r^2}-g^{\circ}\left(r,\theta\right)+w\left(r,\theta\right)-\zeta \ ,
\end{equation}
where $w(r,\theta)$ is the total displacement field and $\zeta=w\left(0,\theta\right)$ is the indentation, {\em i.e.} the displacement at the center.
For light loads and within the apparent contact region, only the top of the pillars can be in contact with the lens, whereas the substrate not covered by pillars will be separated from the lens by a gap.
In addition, one may notice that, within the contact region, the top of each pillar may be either in total or partial contact.
Whatever the contact situation of each pillar, at the local contact regions, one has $\Delta\left(r,\theta\right)=0$ and $g^{\circ}\left(r,\theta\right)=h$, and thus the total displacement field $w\left(r,\theta\right)$ is related to the total indentation $\zeta$ as follows:
\begin{equation}
w\left(r,\theta\right)=\zeta-\left[R-\sqrt{R^2-r^2}\right] \ .
\label{Eq:TotDisp}
\end{equation}


\section{Discrete contact}
\label{Sec:Contact}

For each body $j$ (lens $j=l$, array of pillars $j=p$ and substrate $j=s$), considering its Young's elastic modulus $E_j$ and Poisson's ratio $\nu_j$, we define the parameter $\gamma_j=\left[1-\nu_j^2\right]/E_j$.
Herein, we will always consider that the pillars and the substrate are made of the same material, thus $E_p=E_s$ and $\nu_p=\nu_s$.
As mentioned above, when these bodies are brought into contact, \textit{i.e.} the lens is approached towards the pillars, which in turn pushes the substrate, they suffer a deformation.
Under these circumstances, recalling that the vertical displacement at the surface of a half-space $u_z$ due to a concentrated point force $\Sigma$ is given by~\cite{Johnson}:
\begin{equation}
u_{z}\left(\Sigma,r\right)=\dfrac{\gamma\Sigma}{\pi r} \ ,
\end{equation}
and applying the principle of linear superposition, the normal displacement of the spherical lens and the substrate ($j=l,s$) reads:
\begin{equation}
w_j\left(r,\theta\right)=\iint_\mathcal{A} u_z\left(\sigma,\varrho\right) \varrho d\varrho d\varphi \ ,
\label{Eq:DCont}
\end{equation}
where $\sigma=\sigma\left(\varrho,\varphi\right)$ is the stress field over the apparent contact area $\mathcal{A}$, which is described by $\varrho$ and $\varphi$, the radial and angular coordinates relative to the point $\left(r,\theta\right)$.
In other words, the displacement $w_j$ at the point $\left(r,\theta\right)$ is the effect of the stress field applied over a region described by the set of relative positions $\left(\varrho,\varphi\right)$.
Due to the discrete nature of the contact, the normal displacement of each surface can be decomposed as follows:
\begin{align}
w_j\left(r,\theta\right) &=\dfrac{\gamma_j}{\pi}\sum_{i=0}^{Q}  W_i\left(r,\theta,r_i\ ,\theta_i\right) \ , &
W_i\left(r,\theta,r_i\ ,\theta_i\right) &=\int_{\varphi_{i1}}^{\varphi_{i2}}\int_{\varrho_{i1}}^{\varrho_{i2}} \sigma\left(\varrho,\varphi\right) d\varrho d\varphi \ ,
\label{Eq:DDisc}
\end{align}
where $W_i$ is the reduced displacement at $\left(r,\theta\right)$ due to the contact with the $i$th pillar.
Additionally, we make emphasis on the fact that the displacement field $W_i$ depends explicitly on the position of the $i$th pillar , since the integration limits $\varrho_{i1}$, $\varrho_{i2}$, $\varphi_{i1}$ and $\varphi_{i2}$ must be determined for each combination of pillar location $\left(r_i\ ,\theta_i\right)$ and point $\left(r,\theta\right)$.
For instance, $\varrho_{i1}$ and $\varrho_{i2}$ are the solutions for $\varrho$ of quadratic equation:
\begin{equation}
\varrho^2-\left[2\rho_i\cos\left(\varphi\right)\right]\varrho+\left[\left(\rho_i\right)^2-\left(d/2\right)^2\right]=0 \ ,
\end{equation}
where $\rho_i=\rho_i\left(r,\theta,r_i,\theta_i\right)$ given in eq.~\ref{Eq:Distnmlk}, whereas $\varphi_{i1}$ and $\varphi_{i2}$ depend on the location of the point $\left(r,\theta\right)$.
On one hand, if $\left(r,\theta\right)$ lies inside the contact region of the $i$th pillar, we have $\varphi_{i1}=0$ and $\varphi_{i2}=\pi$, and,  on the other hand, if $\left(r,\theta\right)$ is located outside the contact region of the $i$th pillar, we find:
\begin{equation}
\varphi_{i1,2}=\pm \arcsin\left(\dfrac{\left[d/2\right]\sin\left(\pi/2\right)}{\rho_i}\right) \ .
\end{equation}

In turn, each pillar is compressed because of its confinement between the lens and the substrate.
Herein, we consider that the compression of the $i$th pillar is approximated by the Hooke's law, which gives a change of height equal to $\delta_i=4hF_i/\left(\pi d^2E_p\right)$.
As well, $F_i$ is the load exerted over the $i$th pillar in the contact region, which reads:
\begin{equation}
F_i\left(r_i,\theta_i\right)=\int_{0}^{2\pi}\int_{0}^{d/2} \sigma\left(\rho_i,\phi_i\right)\rho_i d\rho_i d\phi_i \ ,
\label{Eq:FCyl}
\end{equation}
whereas, in this situation, $\rho_i$ and $\phi_i$ represent the radial and angular coordinates at the surface of the $i$th pillar, relative to the pillar center $\left(r_i,\theta_i\right)$ and restricted to the ranges $\rho_i\in\left[0,d/2\right]$ and $\phi_i\in\left[0,2\pi\right]$.
In order to transform the stress field over the entire domain $\sigma\left(r,\theta\right)$ into $\sigma\left(r_i,\theta_i,\rho_i,\phi_i\right)$, one should express $r$ and $\theta$ for a given relative position $\left(\rho_i,\phi_i\right)$ and a pillar $\left(r_i,\theta_i\right)$ using the following expressions:
\begin{align}
r &=\sqrt{\rho_i^2+r_i^2+2\rho_i r_i \cos\left(\phi_i-\theta_i\right)} \ , &
\theta &=\arcsin\left(\dfrac{\rho_i\sin\left(\pi-\phi_i\right)+r_i\sin\left(\theta_i\right)}{r}\right) \ .
\label{Eq:CylRel}
\end{align}

Additionally, the total load $F_T$ reads:
\begin{equation}
F_T=\iint_\mathcal{A} \sigma\left(r,\theta\right) r dr d\theta \ ,
\label{Eq:FCont}
\end{equation}
which, accounting with the discrete nature of the contact, can be obtained by adding the contribution of each pillar as follows:
\begin{equation}
F_T=\sum_{i=0}^{Q}F_i\left(r_i,\theta_i\right) \ .
\label{Eq:FDisc}
\end{equation}

Since, the total displacement field is also composed of:
\begin{equation}
w\left(r,\theta\right)=w_l\left(r,\theta\right)+w_s\left(r,\theta\right)+\delta_i g^{\circ}\left(r,\theta\right)/h \ ,
\end{equation}
which gathers the displacement contributions given in eq.~\eqref{Eq:DDisc} and the Hookean pillar compression $\delta_i$, imposing a total indentation, given by eq.\eqref{Eq:TotDisp} and evaluated at the top of the pillars where $g^{\circ}\left(r,\theta\right)=h$, allows one to recover the stress field $\sigma$ inside the contact region.
Herein, $\sigma$ is computed numerically using a superposition method of discrete pressure elements, for which a segmentation of each contact region (the top of each pillar) into square subregions of constant constraint is performed.
Details are given in the supplementary material \cite{SM}.
A similar approach has been introduced and validated in previous studies \cite{Guidoni2010,Jagota2007}, for which a Hookean hypothesis and a uniform pressure has been applied to each pillar.
Even though our methodology retains the Hookean behavior for each pillar, it allows the appearance of non-axysimmetric pressure fields for the non-centered pillars ($i>0$), thanks to the discretization of each contact region.
This procedure provides a more realistic approach for several cases, including small total loads $F_T$ and small surface fractions $\Phi$, but also the effect of the finite size of the contact area, for any load and any surface fraction.
The corresponding results are presented in the following section.

\section{Results}
\label{Sec:Results}

\begin{table}
\begin{center}
    \begin{tabular}{|c|c|c|}
    \hline
    Parameter & Values & Units \\ \hline
    $R$ & 2 & mm \\ 
    $h$ & 2 & $\mu$m \\ 
    $d$ & 4 & $\mu$m \\ 
    $\Phi$ & $\left[2.3\times10^{-3}, 8.2\times10^{-1}\right]$ & 1 \\ 
    $\zeta$ & $\left[0.005,0.5\right]$ & $\mu$m \\ 
    $E_l$ & $1.6\times10^6$ & Pa \\ 
    $E_p=E_s$ & $\left[10^3,10^{12}\right]$ & Pa \\ 
    $\nu_l$ & 0.5 & 1 \\ 
    $\nu_p=\nu_s$ & 0.5 & 1 \\ \hline
    \end{tabular}
\end{center}
\caption{The values of the parameters used to obtain the presented results.}
\label{Tab:Param}
\end{table}

\begin{figure}
\centering
\subfigure[{$\Phi=0.03$}]{
\centering
\includegraphics[width=0.45\textwidth]{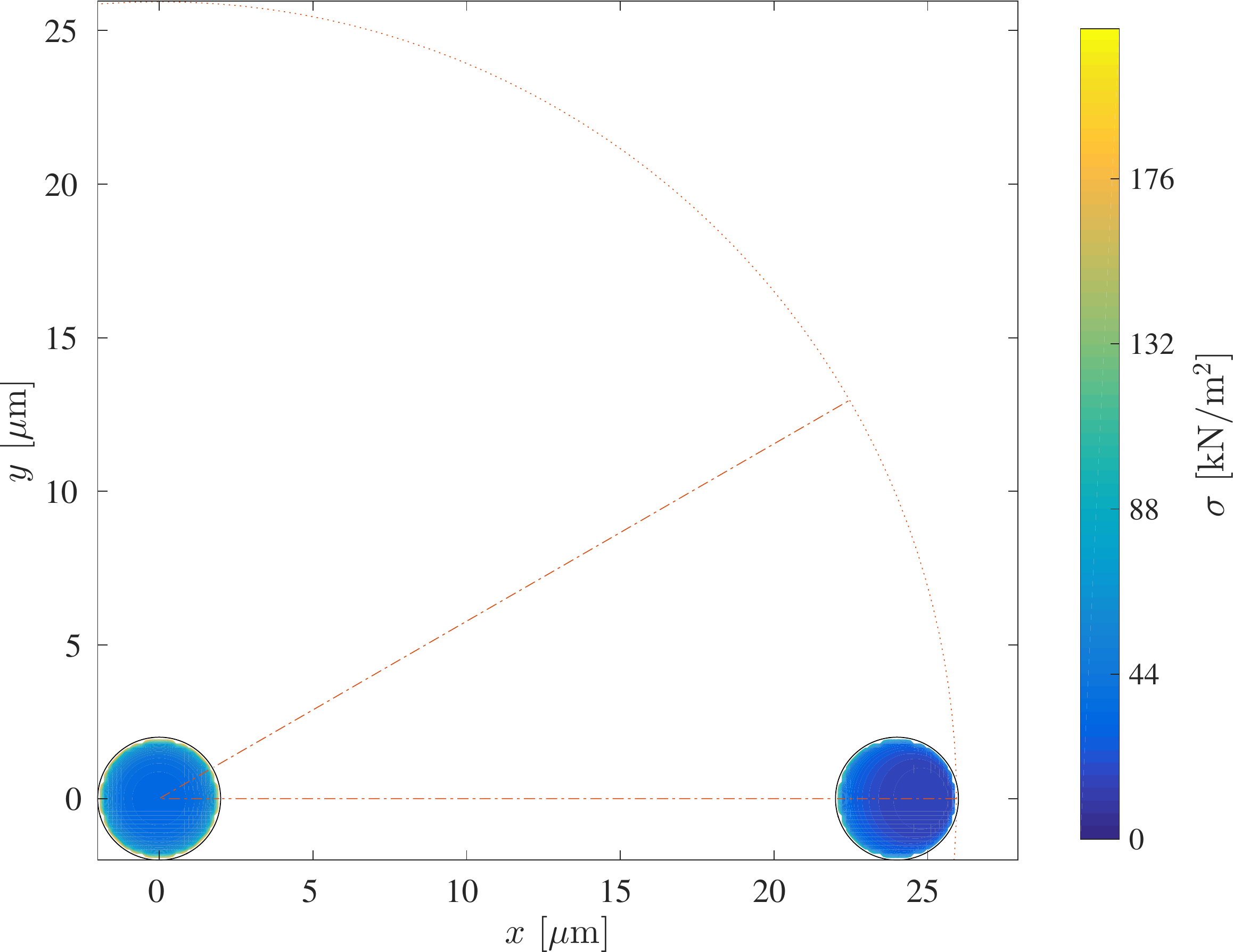}
\hspace*{5mm}
\includegraphics[width=0.35\textwidth]{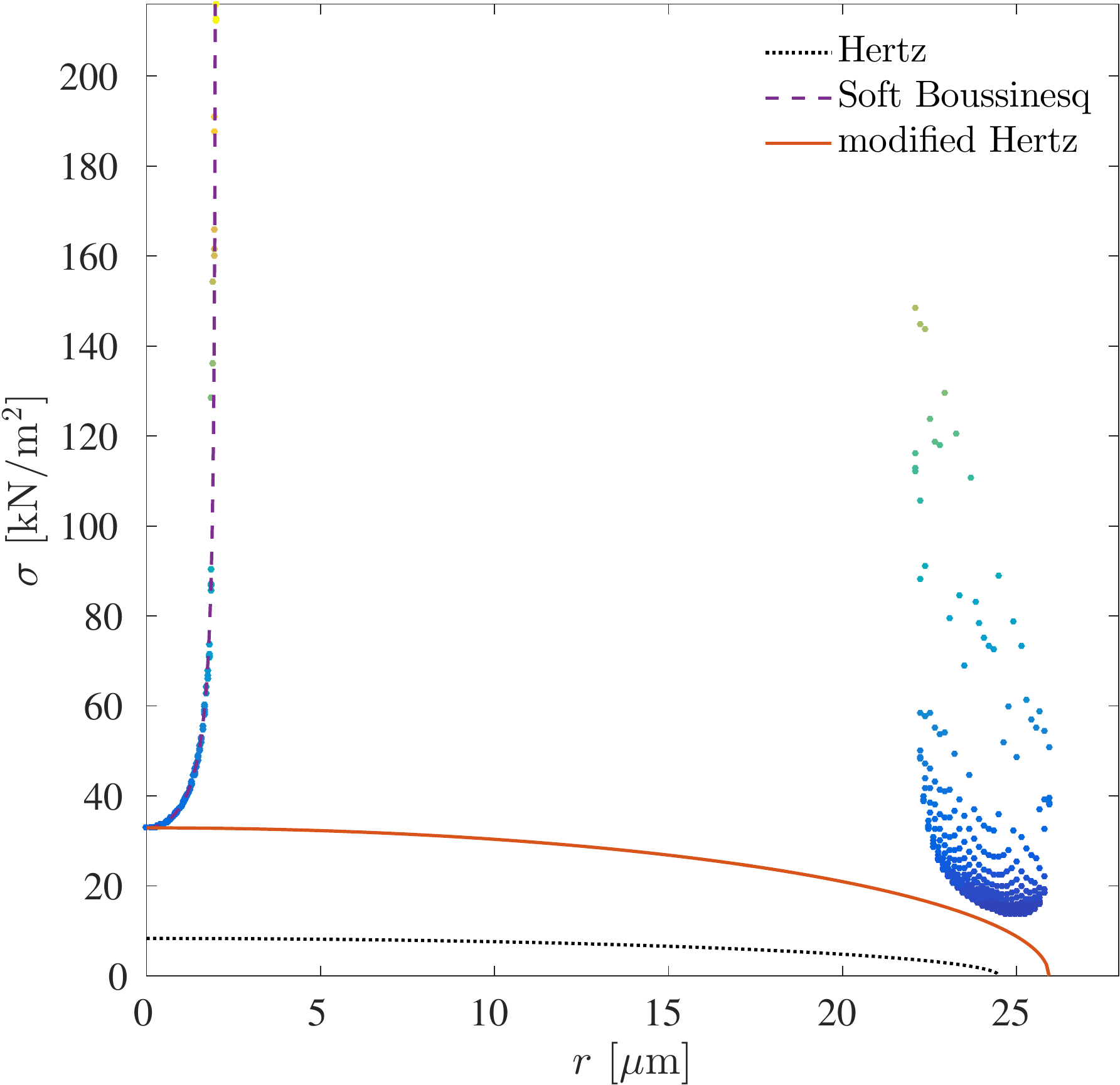}
\label{Fig:Sig2}
}
\\
\subfigure[{$\Phi=0.82$}]{
\centering
\includegraphics[width=0.45\textwidth]{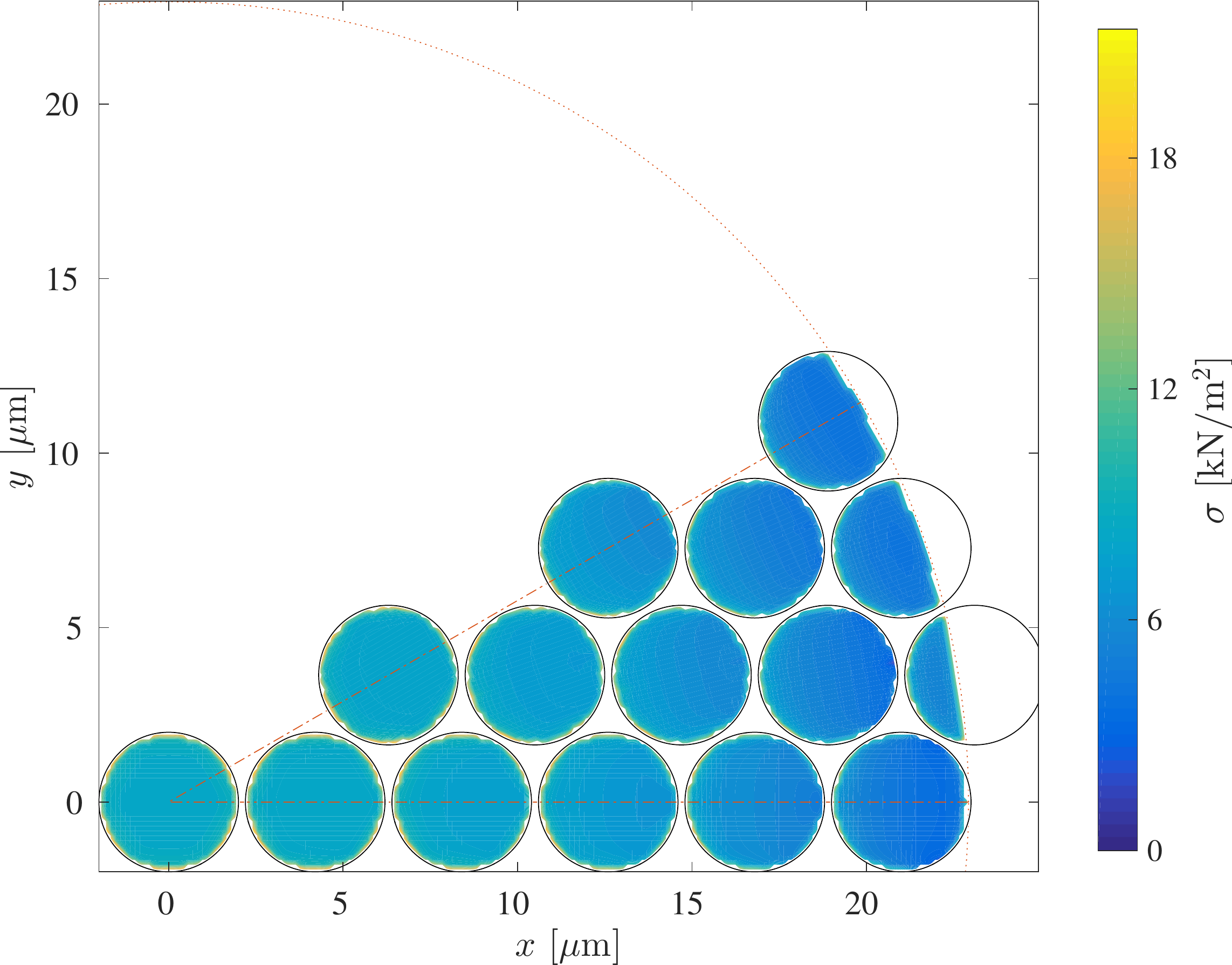}
\hspace*{5mm}
\includegraphics[width=0.35\textwidth]{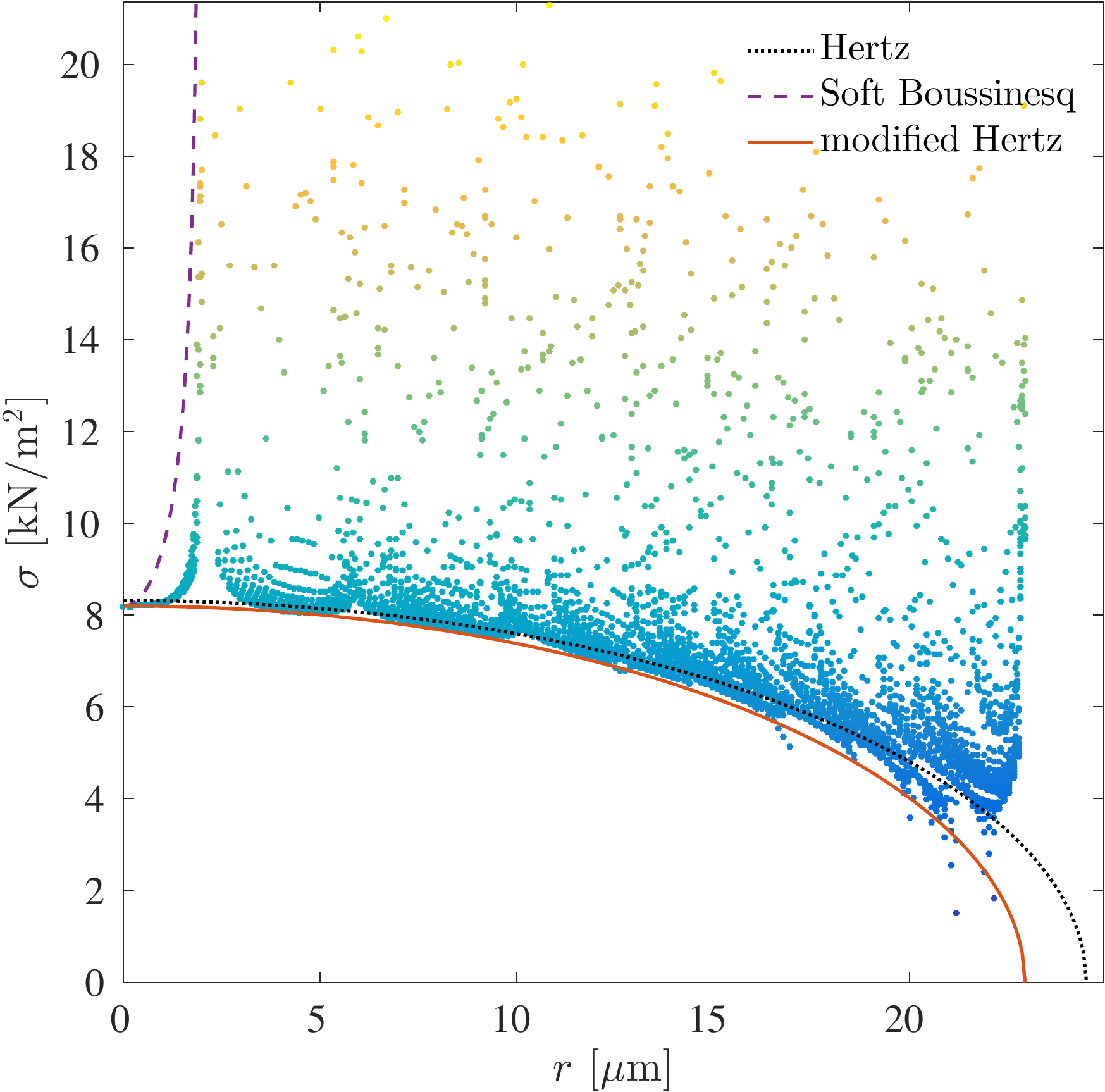}
\label{Fig:Sig4}
}
\caption{Stress field at the contact region $\sigma$ within the primitive cell of the system delimited by the discontinuous lines [\textcolor{red}{- $\cdot$ -}], obtained for $E_p=E_s=1.6\times10^{6}$ Pa, $\zeta=0.3$ $\mu$m, different values of $\Phi$ -- \subref{Fig:Sig2} $\Phi=0.03$ and \subref{Fig:Sig4} $\Phi=0.82$ -- and the remaining parameter values given in Table~(\ref{Tab:Param}). The dotted curves [\textcolor{red}{$\cdots$}] delimitate the computed contact region for each case.
Figures on the left-hand side show a 2D representation in the plane $xy$, whereas figures on the right-hand side show a 1D representation in the axis $r=\sqrt{x^2+y^2}$. In the $\sigma$ \textit{vs} $r$ representation, curves indicating the Hertz and soft-flat-punch stress fields, given respectively by eqs.\eqref{Eq:SFHertz} and \eqref{Eq:SFSFP}, and a modified Hertz-like behavior, for which the stress $\sigma_0$ at $r=0$ and the apparent contact radius $a$ were obtained from the numerical simulations, are also depicted.}
\label{Fig:SigField}
\end{figure}

In Fig.~\ref{Fig:Sig2}, the contact region and the stress field for a small surface fraction $\Phi=0.03$ is depicted for $E_p=E_s=1.6\times10^{6}$ Pa, $\zeta=0.3$ $\mu$m, two particular indentations and all other parameter values given in Table~(\ref{Tab:Param}).
The contact radius is approximately equal to $a\approx e+d/2=26\mu$m.
For the pillar at the center of the contact region, it turns out that the stress field is equivalent to a soft-flat-punch (SFP) stress field. The stress field is axissymetric, with a large gradient,\textit{i.e.} small $\sigma$ around the central zone $r\simeq 0$ and very large $\sigma$ near near the edge of the pillar $r\lesssim d/2$. We clearly observe an increase of an order of magnitude when going from the center to the border of this pillar.
Also, within the primitive cell, a second pillar is in contact, subjected to a stress field which is only symmetric with respect to the plane $\theta=0$.
This field is eccentric with respect to the center of the pillar, leading to slightly oblate iso-stress contour lines, which recover the centric and circular shape as the border of the pillar is reached.
The increase of stress magnitude, from the center to the border of the pillar, is very similar to the one observed for the central pillar.
The stress $\sigma$ at the central region of both pillars seems to follow a Hertzian-like behavior, a square root profile calculated with the stress $\sigma_0$ at $r=0$ and the contact radius $a$ obtained from the simulations.
As expected, relatively high magnitudes of the ``discrete'' stress field are observed in this case, since the total load is concentrated over 7 pillars, considering the geometry of the system. For comparison, the Hertzian stress profile is plotted in the figure.

For the same indentation and Young's modulus, the contact region and the stress field for a large surface fraction $\Phi=0.82$ is illustrated in Fig.~\ref{Fig:Sig4}.
The contact radius is approximately equal to $a\approx 5e+d/2=23\mu$m, slightly smaller than the radius of the small $\Phi=0.03$ case.
The behavior of $\sigma$ at the central pillar deviates strongly from the SFP profile, with a nearly constant stress magnitude at the center region $0\leq r\leq d/4$, that diverges rapidly as the border is reached (a stress profile similar to the ones found by Gecit~\cite{Gecit1986}).
Nevertheless, the stress profile remains axisymmetric and its magnitude is relatively small over the entire lid of the pillar.
For the other pillars, which are not at $r=0$, the local stress field is only symmetric with respect to the corresponding plane $\theta=\theta_i$ of the $i$th pillar.
Oblate iso-stress contour zones, nearly crescent-like shapes, are found, mainly for the pillars near the border of the contact region.
Once more, considering $\sigma$ at the central region of each pillar, a Hertzian stress field is observed, with the usual square root profile computed with a stress $\sigma_0$ at $r=0$, very close to the true Hertz value, and a contact radius $a$ that is significantly smaller than the true Hertz contact radius.

It is important to notice that, while a constant indentation $\zeta$ is kept for the cases presented in Fig.~\ref{Fig:SigField}, a reduction of $\sigma_0$, of one order of magnitude (from 33 to 8 kPa) when $\Phi$ is increased (from 0.03 to 0.82), is observed.
As well, the same increase of $\Phi$ produces a reduction of the apparent contact region and the contact radius $a$ (from 26 to 23 $\mu$m), but the real contact area given by the number of pillars in contact has increased, even though the total load $F_T$ has risen (from 3 to 10 $\mu$N).

\begin{figure}
\centering
\subfigure[]{
\centering
\includegraphics[width=0.45\textwidth,trim=0cm 0cm 0cm 0cm,clip=true]{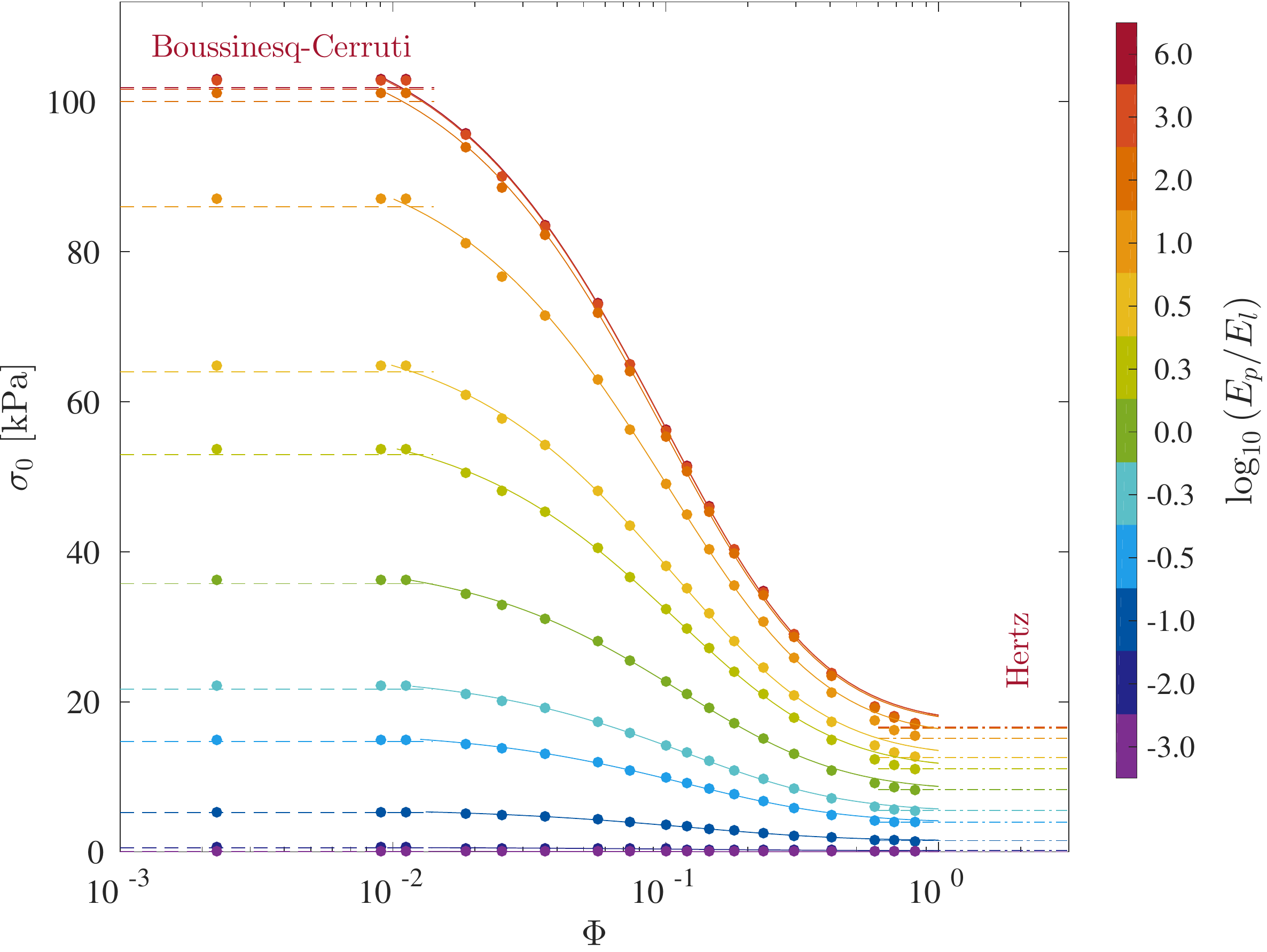}
\label{Fig:Sig0Phi}
}
\hspace*{5mm}
\subfigure[]{
\centering
\includegraphics[width=0.41\textwidth,trim=0cm 0cm 0cm 0cm,clip=true]{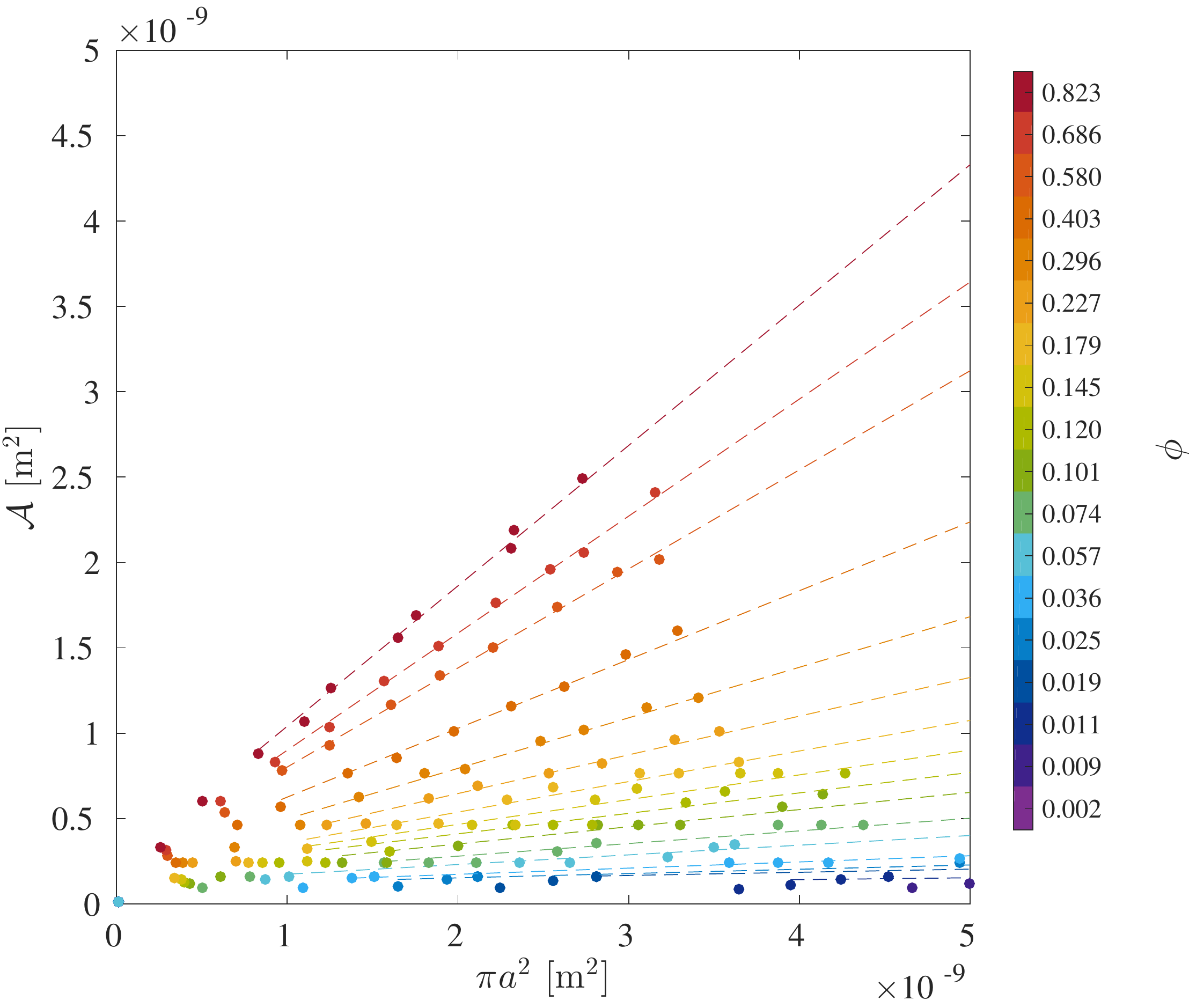}
\label{Fig:APhi}
}
\caption{\subref{Fig:Sig0Phi} Stress at the center of the contact region $\sigma_0$, for $\zeta=0.3$ $\mu$m and the remaining parameters given in Table~(\ref{Tab:Param}). Continuous color lines (from [\textbf{\textcolor{red}{---}}] hard to [\textbf{\textcolor{violet}{---}}] soft pillars-substrate materials) were obtained from eq.~\eqref{Eq:MasterSig0Phi} and the corresponding fit of $\Phi_H\in\left[0.306,0.38\right]$ and $\Phi_S\in\left[0.0094,0.0135\right]$ -- both from hard to soft materials --, for each set of parameters. The Hertzian and soft-flat-punch (SFP) values were computed with eqs.~\eqref{Eq:S0Hertz} and \eqref{Eq:S0SFP}, respectively. 
\subref{Fig:APhi} Real contact area $\mathcal{A}$ as functions of the apparent contact area $\pi a^2$, for $E_p/E_l=1$ and the remaining parameters given in Table~(\ref{Tab:Param}).
Dashed lines (from [\textbf{\textcolor{red}{- -}}] high to [\textbf{\textcolor{violet}{- -}}] low surface fraction $\Phi$) indicate the linear dependence $\mathcal{A}=\Phi\left[\pi a^2\right]$, expected to be the asymptotic behavior for relatively large indentations $\zeta$ and, thus, large apparent contact radius $a$.}
\label{Fig:Sig0APhi}
\end{figure}

In Fig.~\ref{Fig:Sig0Phi}, the dependence of the stress at the center of the contact region $\sigma_0$ on the surface fraction $\Phi$, for an indentation of $\zeta=0.3$ $\mu$m and the remaining parameter values given in Table~(\ref{Tab:Param}), is depicted.
Going from right to left, when the surface fraction is $\Phi=1$, the Hertzian value of $\sigma_0$ is found.
As $\Phi$ decreases, the total load distribution over a smaller number of pillars induces a slightly larger apparent contact region, but it provokes a stress concentration, represented by an increase of $\sigma_0$, due to the reduction of the real contact area and the number of pillars in contact.
If we continue to diminish $\Phi$, a monotonic growth of $\sigma_0$ is observed until it saturates to a plateau.
This saturation, which corresponds to a SFP behavior, is completely attained when the surface fraction $\Phi$ becomes small enough to leave just a single pillar within the contact region.
In Fig.~\ref{Fig:Sig0Phi}, the Hertz-to-SFP transition is presented for different values of the ratio $E_p/E_l$, for the same indentation $\zeta$.
A reduction of the system stiffness $E_p/E_l$ enhances a decrease of the $\sigma_0$ stress levels observed for both limits, Hertz and SFP.
Despite this vertical shift, the general trend remains unchanged whatever the value of the ratio $E_p/E_l$.

In Fig.~\ref{Fig:APhi}, the real contact area $\mathcal{A}$ is presented in terms of the apparent contact area $\pi a^2$, for a Young's moduli ratio $E_p/E_l=1$ and all other parameters given in Table~(\ref{Tab:Param}).
The following description remains qualitatively valid for any value of the ratio $E_p/E_l$, presented in this work.
Note that data corresponding to very small values of the surface fraction $\Phi<0.01$ are included but not visible, because they are superimposed near the origin since the real and apparent contact areas are equal and correspond to the top surface of a pillar $\mathcal{A}=\pi a^2=\pi d^2/4\sim 5\times 10^{-11}$ $m^2$, which is two orders of magnitude smaller than most of the presented data.
Describing each curve from right to left, when the indentation $\zeta$ is small, both the apparent $\pi a^2$ and the real $\mathcal{A}$ contact areas are small, with $\mathcal{A}<\pi a^2$.
As the applied total load $F_T$, the indentation $\zeta$ and the apparent contact area $\pi a^2$ are increased, $\mathcal{A}$ grows monotonically and approaches the trend $\Phi\left[\pi a^2\right]$, for all the values of $\Phi$.
For the highest values of the surface fraction $\Phi$, the real contact area $\mathcal{A}$ approaches the Hertzian value $\mathcal{A}_H$, which is a circle of radius $a=\sqrt{R\zeta}$.
As we diminish the surface fraction $\Phi$, the number of pillars in contact is reduced, provoking a drop of $\mathcal{A}$.
At a certain value of $\Phi$, which is smaller for the highest Young's moduli ratios $E_p/E_l$, the real and apparent contact areas suddenly take the value of the SFP contact, \textit{i.e.} the area of the top of a single pillar $\mathcal{A}_S=\pi d^2/4$.

\section{Asymptotic cases}
\label{Sec:Asymptotic}

There are special situations for which the size of the contact region, the contact radius $a$, can be predicted analytically.
They correspond to the two limiting cases of the surface fraction, which are depicted in Fig.~\ref{Fig:Asymptotic}: $\Phi\rightarrow 1$ for Hertzian and $\Phi\rightarrow 0$ for soft-flat-punch contact.
Both behaviors are briefly described in the following subsections.

\subsection{Hertzian contact}
\label{Sec:HertzCont}

Hertzian contact~\cite{Johnson} occurs either when $\Phi\rightarrow 1$ or when the contact radius is $a\ll d/2$. 
Under these conditions, the problem simplifies to an axisymmetric geometry for which the pillar array becomes a layer of finite thickness~\cite{Li1997}.
Considering infinite friction between the two bodies, the well-known stress field for Hertzian contact reads:
\begin{equation}
\sigma\left(r\right)=\sigma_{0H}\sqrt{1-\left(\dfrac{r}{a}\right)^2} \ ,
\label{Eq:SFHertz}
\end{equation}
where the contact radius is $a=\sqrt{\zeta R}$, and the stress magnitude $\sigma_{0H}$, perceived at the center of the contact region $r=0$, is computed from the total indentation $\zeta$ as follows:
\begin{equation}
\sigma_{0H}=\dfrac{2 }{\pi \left[\gamma_l+\gamma_s\right]}\sqrt{\dfrac{\zeta}{R}} \ .
\label{Eq:S0Hertz}
\end{equation}
Therefore, since the pillars/substrate had become a single entity, $\sigma_{0H}$  is equally sensitive to a change in stiffness of any component of the system (lens and compound substrate).
In addition, as it is explicitly indicated in eq.\eqref{Eq:S0Hertz}, a larger value of the stress $\sigma_{0H}$ is expected when an important indentation $\zeta$ is imposed.
It is also well-known that the total force is $F_T=\left(2\pi/3\right) R \zeta \sigma_{0H}$ .

\subsection{Soft-flat-punch contact}
\label{Sec:SFPCont}

When $\Phi\rightarrow 0$, a soft-flat-punch elastic contact is observed, since only one pillar is in contact with the lens.
In the case where $E_p\gg E_l$, we can consider that the pillars/substrate system is rigid $\delta_0=w_s\left(r,\theta\right)=0$ and, if the top of the pillar is completely in contact with the lens $a=d/2$, the problem is described by a uniform normal displacement generated with a rigid flat punch~\cite{Sneddon1945,Johnson,Maugis}.
The solution for the aforementioned problem corresponds to the superposition of the SFP solution, within a circular region at which uniform displacement is imposed.
This situation, for which the well-known stress field reads:
\begin{equation}
\sigma\left(r\right)=\sigma_{0S}\left(\scriptstyle{E_p\rightarrow\infty}\right)\left[1-\left(\dfrac{2r}{d}\right)^2\right]^{-1/2} \ ,
\label{Eq:SFSFP}
\end{equation}
is valid for the lens-pillar-substrate system only for a total indentation $\zeta$ that is smaller than the pillar height $h$.
Herein, the stress magnitude $\sigma_{0S}\left(\scriptstyle{E_p\rightarrow\infty}\right)$ at $r=0$, the center of the pillar, is computed from the total indentation $\zeta$ as follows:
\begin{equation}
\sigma_{0S}\left(\scriptstyle{E_p\rightarrow\infty}\right)=\dfrac{2\zeta}{\pi d \gamma_f} \ .
\label{Eq:S0SFPInf}
\end{equation}

In contrast, the case for which $E_p\simeq E_l$ does not presents an analytical model, and must be studied numerically~\cite{Gecit1986}.
Here-in, we will name this case ``soft-flat-punch'' contact problem, for which we propose the following simple model.
Considering that the total indentation $\zeta$ equals the addition $w_l\left(0,\theta\right)+w_s\left(0,\theta\right)+\delta_0$, and the expressions $F_0=\pi d^2\sigma_{0S}/2$ and $w_{l,s}\left(0,\theta\right)=\pi d \gamma_{l,s}\sigma_{0S}/2$,which come from the rigid-flat-punch solution (the latter by making $\zeta=w_{l,s}\left(0,\theta\right)$ and $\sigma_{0S}\left(\scriptstyle{E_p\rightarrow\infty}\right)=\sigma_{0S}$ in eq.\eqref{Eq:S0SFPInf}), and that the compression of the center pillar is $\delta_0=4hF_0/\left(\pi d^2E_p\right)$, we can write:
\begin{equation}
\zeta\approx\dfrac{\pi d}{2}\left[\gamma_l+\gamma_s+\dfrac{4h}{\pi d E_p}\right]\sigma_{0S} \ .
\end{equation}
This relationship resembles a combination of springs in series, which configuration is described by an effective elastic parameter $\gamma^{\ast}$ constructed as follows:
\begin{equation}
\gamma^{\ast}\approx\dfrac{1-\nu_f^2}{E_l}\left\{1+\left[\dfrac{1-\nu_s^2}{1-\nu_l^2}\right]\dfrac{E_l}{E_s}+\dfrac{4h E_l}{\pi d \left[1-\nu_f^2\right]E_p}\right\} \ .
\end{equation}
In particular, when $E_p\gg E_l$ and recalling that $E_s=E_p$, which corresponds to a rigid flat punch and an elastic half-space since the pillar/substrate can be taken as a rigid body, and also considering that $\nu_l\sim\nu_p$ and $h\sim d$, leads us to recover $\gamma^{\ast}\left(\scriptstyle{E_p\rightarrow\infty}\right)=\gamma_l$.

\begin{figure}
\centering
\includegraphics[width=0.45\textwidth]{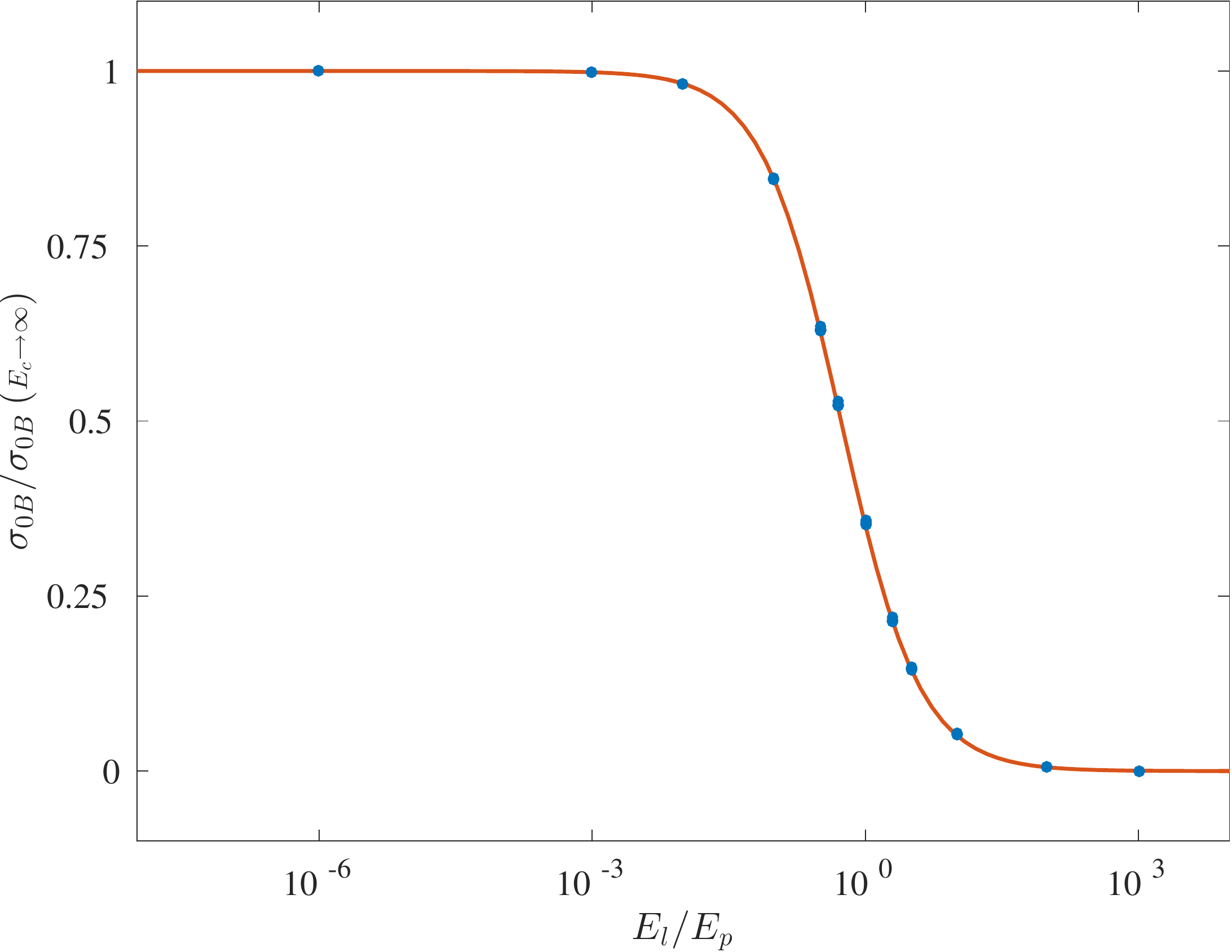}
\caption{Stress at the center of the pillar $\sigma_{0S}$, for the soft-flat-punch contact, as a function of the ratio of Elastic moduli $E_l/E_p$, for $\Phi=2.3\times10^{-3}$ and the remaining values given in Table~(\ref{Tab:Param}). The behavior described by the continuous curve [\textcolor{red}{---}] was computed with eq.\eqref{Eq:S0BvsE}.}
\label{Fig:Sig0SFP}
\end{figure}

For the contact between the lens and a soft pillar, employing the effective elastic constant, we can define the stress at $r=0$ as:
\begin{equation}
\sigma_{0S}=\dfrac{2\zeta}{\pi d \gamma^{\ast}} \ .
\label{Eq:S0SFP}
\end{equation}
In addition, for a constant total indentation $\zeta$ and considering $E_s=E_p$, the comparison between the previous definition and eq.\eqref{Eq:S0SFPInf} yields the following expression for $\sigma_{0S}$ in terms of the ratio of Elastic moduli $E_l/E_p$:
\begin{equation}
\sigma_{0S}=\sigma_{0S}\left(\scriptstyle{E_p\rightarrow\infty}\right)\left\{1+\left(\dfrac{1-\nu_p^2}{1-\nu_l^2}+\dfrac{4h}{\pi d \left[1-\nu_f^2\right]}\right)\dfrac{E_l}{E_p}\right\}^{-1} \ .
\label{Eq:S0BvsE}
\end{equation}
In Fig.\ref{Fig:Sig0SFP}, this relationship is shown, which sigmoid behavior shows an excellent agreement with the numerical results.
For $E_l/E_p<10^{-3}$, a saturation towards the Boussines-Cerruti solution is observed $\sigma_{0S}\simeq\sigma_{0S}\left(\scriptstyle{E_p\rightarrow\infty}\right)$.
As the ratio $E_l/E_p$ increases, the value of $\sigma_{0S}$ falls down, reaching a new plateau of level $\sigma_{0S}\simeq 0$, when $E_l/E_p>10^{3}$.
Therefore, in the latter regime, for which the lens is extremely soft compared to the pillar/substrate stiffness, the stress $\sigma_{0S}$ is negligible.

\section{Discussion}
\label{Sec:BHtransition}

\begin{figure}
\centering
\subfigure[]{
\centering
\includegraphics[width=0.45\textwidth,trim=0cm 0cm 0cm 0cm,clip=true]{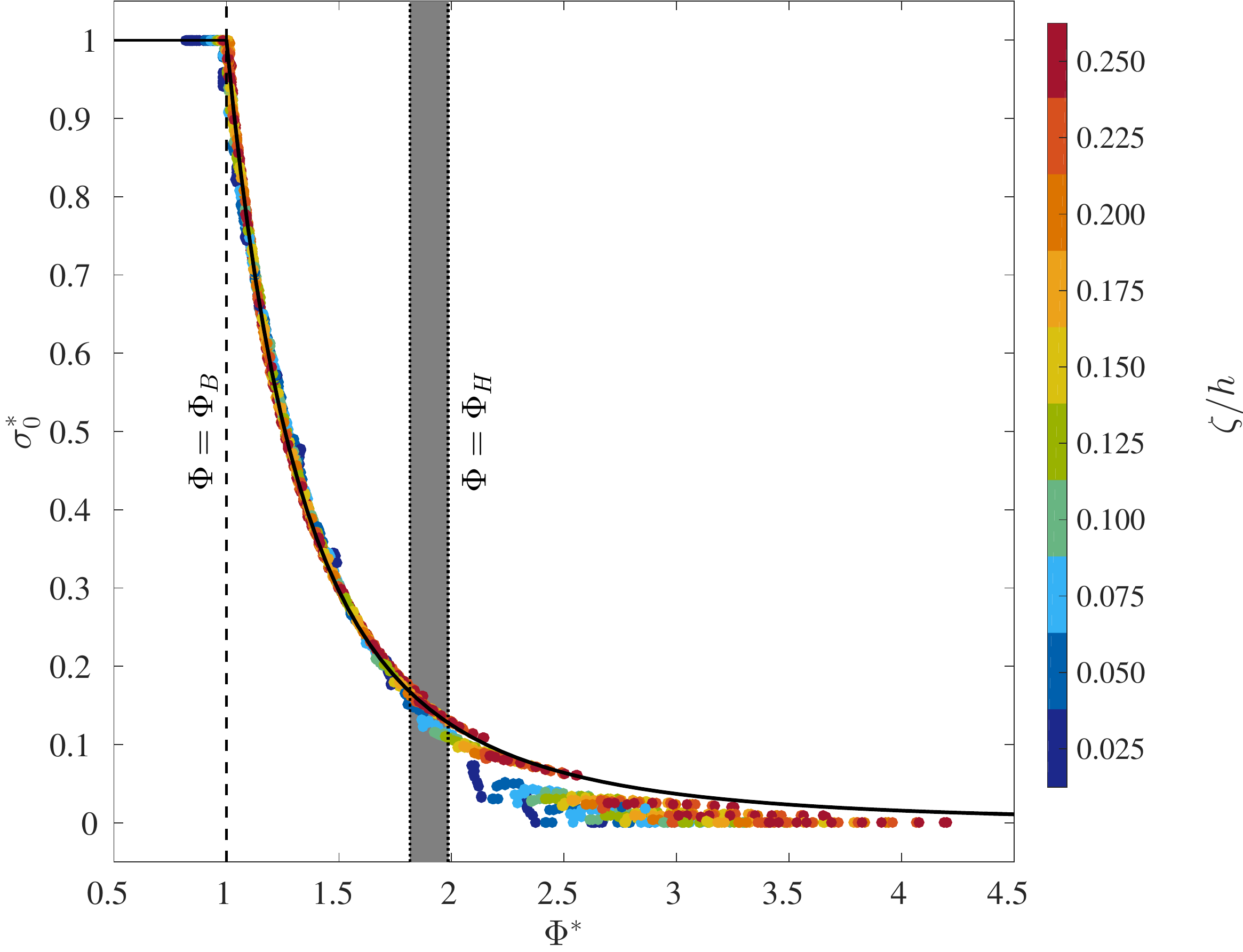}
\label{Fig:Sig0PhiDL}
}
\hspace*{5mm}
\subfigure[]{
\centering
\includegraphics[width=0.45\textwidth,trim=0cm 0cm 0cm 0cm,clip=true]{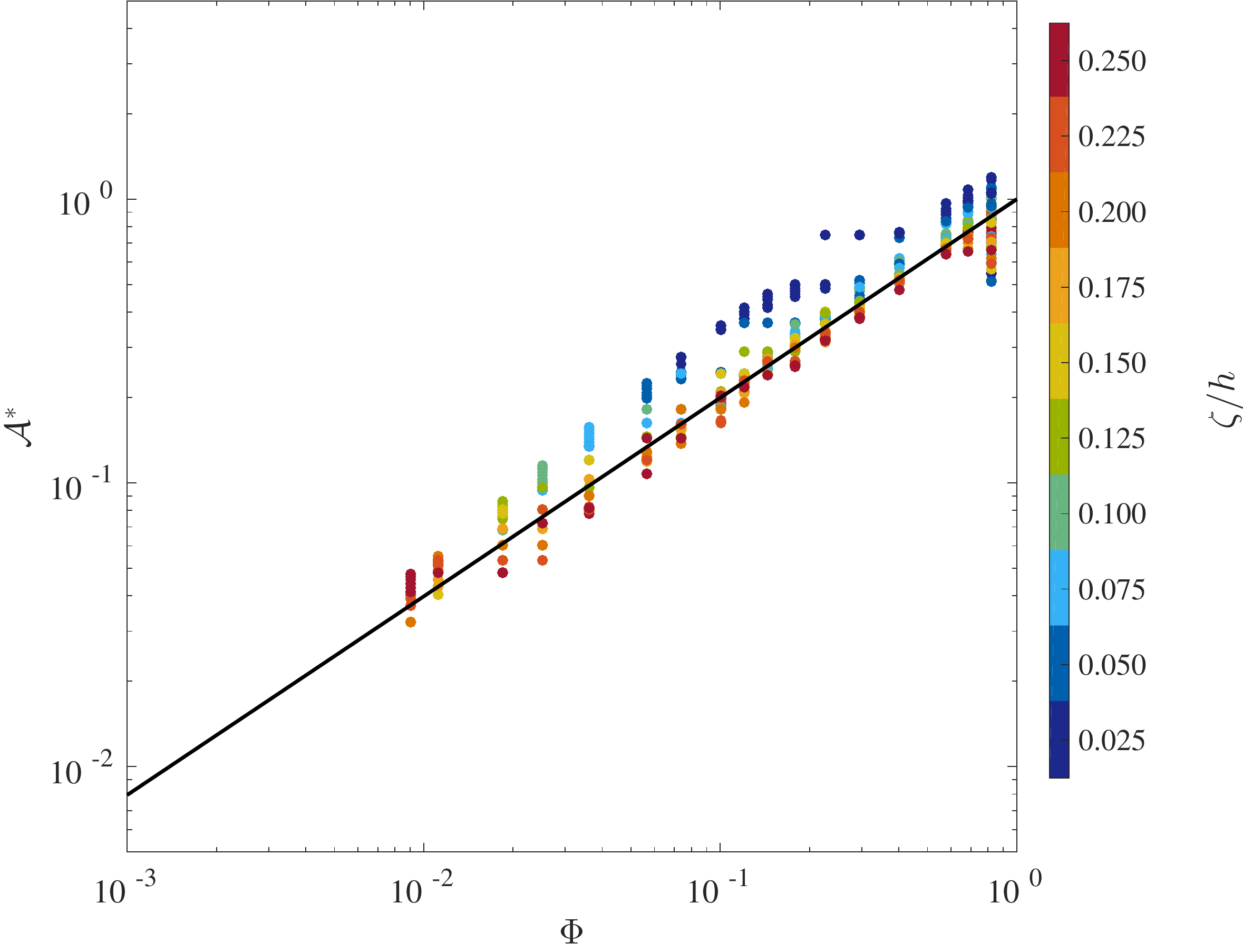}
\label{Fig:APhiDL}
}
\caption{\subref{Fig:Sig0PhiDL} Reduced stress at the center of the pillar $\sigma_0^{\ast}$ as function of the reduced surface fraction $\Phi^{\ast}$ and \subref{Fig:APhiDL} reduced real contact area $\mathcal{A}^{\ast}$ as a function of the surface fraction $\Phi$, in the range $0.05\leq\zeta/\leq0.5$ $\mu$m and for the parameters given in Table~(\ref{Tab:Param}). The trends depicted by the continuous lines [\textbf{\textcolor{black}{---}}], corresponds to the empiric relationships given by eq.~\eqref{Eq:MasterSig0Phi} for \subref{Fig:Sig0PhiDL}, and eq.~\eqref{Eq:MasterAPhi} for \subref{Fig:APhiDL}.}
\label{Fig:Sig0APhiDL}
\end{figure}

\begin{figure}
\centering
\subfigure[]{
\centering
\includegraphics[width=0.45\textwidth]{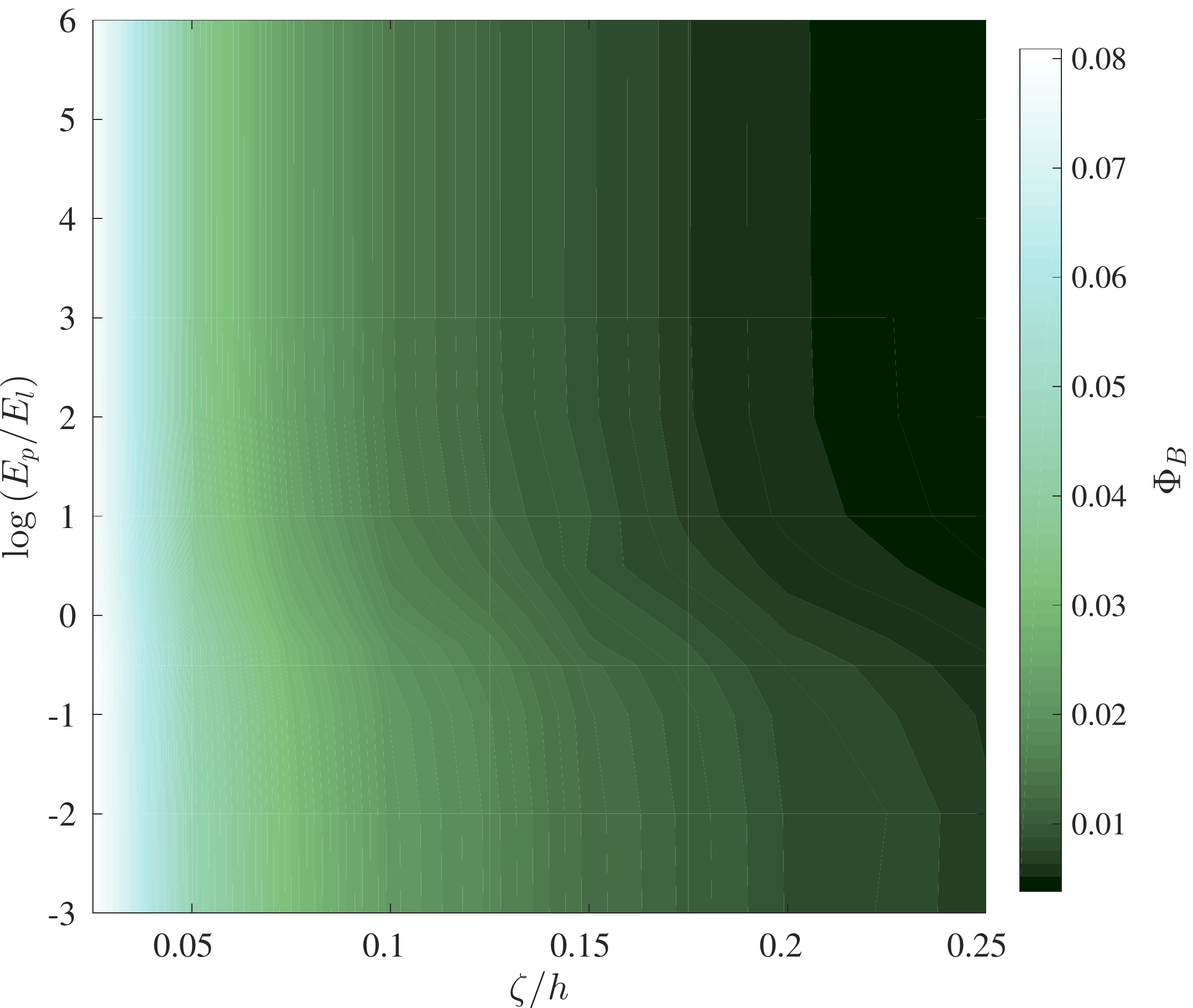}
\label{Fig:PhiB}
}
\hspace{5mm}
\subfigure[ ]{
\centering
\includegraphics[width=0.45\textwidth]{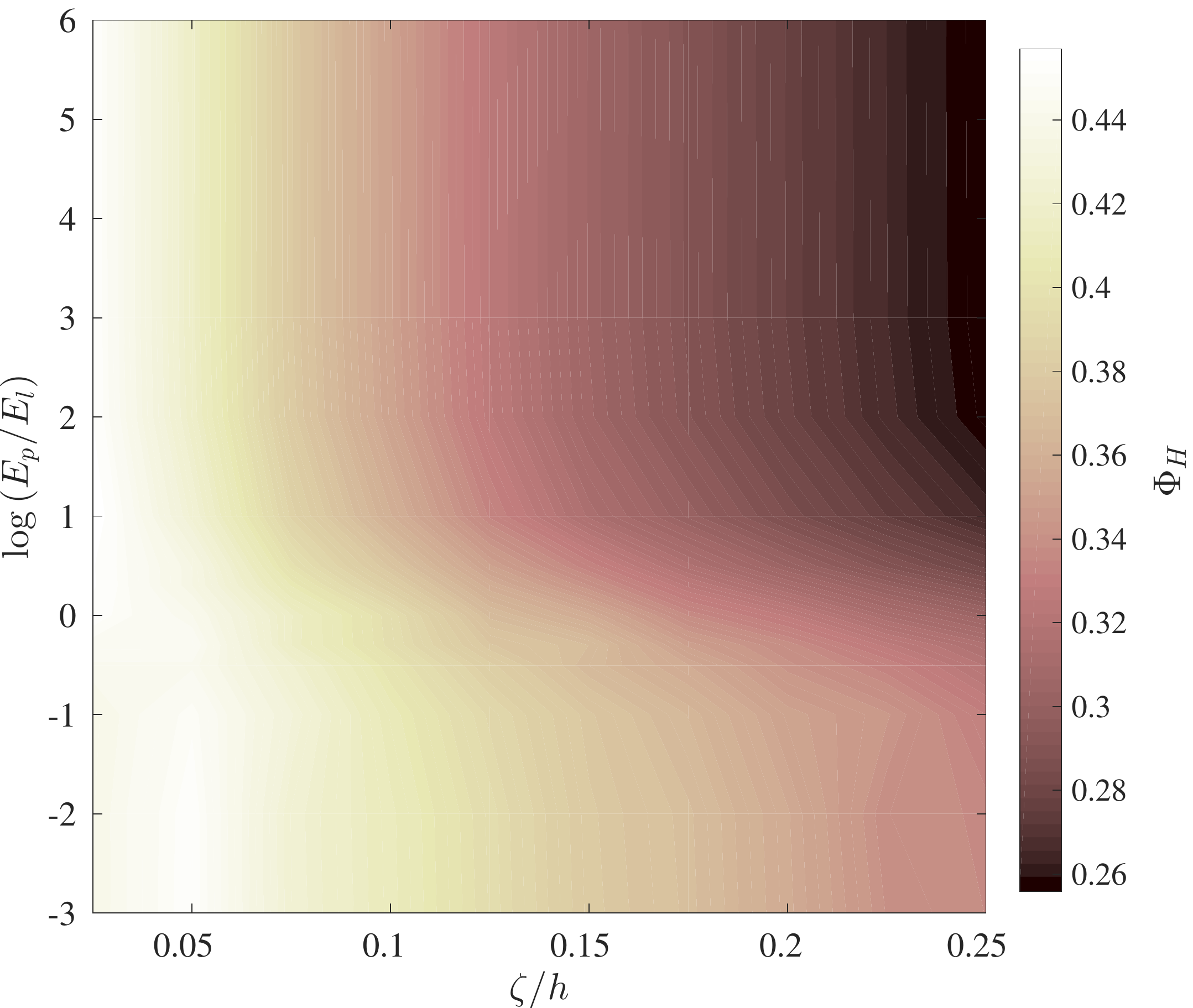}
\label{Fig:PhiH}
}
\caption{\subref{Fig:PhiB} $\Phi_S$ and \subref{Fig:PhiH} $\Phi_H$ as functions of the reduced indentation $\zeta/h$, in the range $0.05\leq\zeta/\leq0.5$ $\mu$m and for the remaining parameters given in Table~(\ref{Tab:Param}).}
\label{Fig:PhiHB}
\end{figure}

Starting with a relatively slight load, provoking the contact of the lens with a single pillar, the soft-flat-punch contact behavior is observed until a second pillar (six pillars at different angular position but the same radial coordinate) is touched by the lens, which is due to an increase of the load and, consequently, the total indentation.
Under this circumstances, from the contact condition $\Delta\left(r,\theta\right)=0$, we find the relationship between the indentation $\zeta$ and the radial position $r_S$, at which a second pillar must be placed in order to be barely touched by the lens:
\begin{equation}
R-\sqrt{R^2-r_S^2}+\left[\dfrac{2}{\Pi}\arcsin\left(\dfrac{d}{2r_S}\right)-1\right]\zeta=0 \ ,
\label{Eq:rSzeta}
\end{equation}
with $\Pi=\pi+4h/\left[d \left(\gamma_g+\gamma_f\right)E_s\right]$.
For a given $\zeta$, we can estimate the radial position $r_S$ at which a second pillar may be reached and, consequently, the corresponding pitch distance $e_S=r_S$ and the SFP threshold surface fraction:
\begin{equation}
\Phi_S=\dfrac{\pi}{2\sqrt{3}}\left(\dfrac{d}{r_S}\right)^2 \ .
\label{Eq:PhiSrS}
\end{equation}
Therefore, when the combination of both parameters, indentation $\zeta$ and surface fraction $\Phi$, is such that $\Phi<\Phi_S$, only one pillar must be in contact with the lens.


Considering that $R\gg r_S\gg d/2$, for relatively large indentations $\zeta\gg \left(3/2\right)^3R\left(d/\Pi\right)^2$, we can approximate the radial position with the expression $r_S\approx \sqrt{2 R \zeta}$, which leads to a threshold surface fraction that scales as $\Phi_S\sim d^2/\left(R \zeta\right)$.

Now, from the opposite situation, we define $\Phi_H$ as another threshold surface fraction, above which a Hertzian contact can be used to approximate the magnitude of the stress field at the center of each pillar.
For high values of $\Phi$, this behavior has been discerned qualitatively from the numerical results, as it was shown in Fig.\ref{Fig:Sig4}.
Nevertheless, for a given indentation and a set of elastic parameters, a quantitative estimation of $\Phi_S$ is only achieved numerically.

Based on this insight, we introduce the reduced variables $\sigma_0^{\ast}$, the stress at the center of the central pillar, and $\Phi^{\ast}$, the reduced surface fraction, define as:
\begin{align}
\sigma_0^{\ast} &=\dfrac{\sigma_0\left(\Phi\right)-\sigma_{0H}}{\sigma_{0S}-\sigma_{0H}} \ , &
\Phi^{\ast} &=\dfrac{\Phi-\Phi_S}{\Phi_H} \ ,
\label{Eq:Sig0PhiRed}
\end{align}
where $\sigma_{0H}$ and $\sigma_{0S}$ are the limiting stresses at $r=0$ for the Hertz and the SFP contact conditions, \textit{i.e.} when $\Phi<\Phi_H$  and $\Phi>\Phi_S$ respectively.
Herein, both stress values $\sigma_{0H}$ and $\sigma_{0S}$ are obtained numerically for $\Phi\sim 1$ and $\Phi\sim 0$, whereas $\Phi_H$  and $\Phi_S$ result from fitting the data of the curves shown in Fig.~\ref{Fig:Sig0Phi} and their equivalents for the different values of indentation $\zeta$ and Young's modulus ratio $E_p/E_l$ given in Table~(\ref{Tab:Param}).

As shown in Fig.~\ref{Fig:Sig0PhiDL}, all the data, including the curves in Fig.~\ref{Fig:Sig0Phi} and the results for different values of $\zeta$, collapse to a master curve when the reduced variables given in eq.\eqref{Eq:Sig0PhiRed} are used.
The following rational function:
\begin{equation}
\sigma_0^{\ast}=\left[1+\Phi^{\ast}\right]^{-3} \ ,
\label{Eq:MasterSig0Phi}
\end{equation}
provides a good qualitative description of the general trend for $\Phi^{\ast}\geq1$, and an excellent quantitative agreement for $\Phi^{\ast}\in\left[0,1-\Phi_S/\Phi_H\right]$.
Therefore, the stress concentration and the growth of $\sigma_0$, perceived during the transition from a Hertzian contact, along a discrete contact region, to a single punch-like contact, behaves as the inverse of the cube of a quantity that measures the separation from the limiting regimes: the difference of the surface fraction from the SFP threshold divided by the Hertz threshold value.
Outside this transition zone, we recover the limiting cases: on one hand, $\Phi^{\ast}=0$ implies that $\sigma_0^{\ast}=1$ and, as a consequence, the SFP behavior $\sigma_0\left(\Phi\right)=\sigma_{0S}$ is expected, whereas on one other hand, $\Phi^{\ast}=\left(1-\Phi_S\right)/\Phi_H$ implies that $\sigma_0^{\ast}\approx 0$ and a Hertzian contact $\sigma_0\left(\Phi\right)\approx \sigma_{0H}$ should represent a good illustration of the discrete contact phenomenon.

The dependencies of $\Phi_S$  and $\Phi_H$ on the indentation $\zeta$ and the Young's modulus ratio $E_p/E_l$ are depicted in Figs.~\ref{Fig:PhiB} and \ref{Fig:PhiH} respectively.
For a fixed value of the ratio $E_p/E_l$, both threshold parameters $\Phi_S$  and $\Phi_H$ follow a similar monotonical decreasing tendency as the reduced indentation $\zeta/h$ increases.
The effect of the ratio $E_p/E_l$ seems to be restricted to a shift towards larger values of both thresholds, strongly for $\Phi_H$ and almost indistinguishably for $\Phi_S$.
It is also important to notice that the values that $\Phi_S$ takes, are an order of magnitude smaller than the values presented by $\Phi_H$. 

For very small indentations $\zeta/h<0.025$, the determination of the values of $\Phi_S$ and $\Phi_H$ becomes delicate, since the Hertzian and SFP stresses at $r=0$ present a very close magnitude $\sigma_{0S}\sim\sigma_{0H}$.
Matching eqs.\eqref{Eq:S0Hertz} and \eqref{Eq:S0SFP}, gives the threshold indentation $\zeta_{T}/h=\left[\gamma^{\ast}/\left(\gamma_l+\gamma_s\right)\right]^2\left[d^2/R h\right]$, which for the parameters given in Table~(\ref{Tab:Param}) varies between $0.004\leq\zeta_T/h\leq0.014$ and acts as an inferior boundary for the swept range of indentations.
Nevertheless, an estimation of both parameters may be provided for $\zeta/h<0.025$.
From the curve of $\sigma^{\ast}$ as a function of $\Phi^{\ast}$ for the last performed fit, \textit{i.e.} for $\zeta/h=0.025$, $\Phi_H$ corresponds to the value of $\Phi$ for which $\sigma^{\ast}_0=0.17$, whereas $\Phi_S$ is to the value of $\Phi$ that makes $\sigma^{\ast}_0=0.92$.
Using this criterion, $\Phi_H$ and $\Phi_S$ can be approximated for $\zeta/h<0.025$.
This results are shown in Fig.~\ref{Fig:Phi0DLInd} and will be discussed in the conclusions section.

We introduce the reduced variable $\mathcal{A}^{\ast}$, the dimensionless real contact area, define as:
\begin{equation}
\mathcal{A}^{\ast}\left(\Phi\right)=\dfrac{\mathcal{A}\left(\Phi\right)-\mathcal{A}_S}{\mathcal{A}_H-\mathcal{A}_S} \ ,
\label{Eq:APhiRed}
\end{equation}
where $\mathcal{A}_H$ and $\mathcal{A}_S$ are the limiting areas for the Hertz and the SFP contact conditions.
Both values are obtained by computing the area of a circular region of radius $a=\sqrt{R\zeta}$ for the Hertzian case, whereas $a=d/2$ for the SFP situation.
In Fig.~\ref{Fig:APhiDL},  all the data, including the curves in Fig.~\ref{Fig:APhi} and the results for different values of $\zeta$, show a similar behavior when the reduced variable, given in eq.\eqref{Eq:APhiRed}, is used in combination with the surface fraction $\Phi$.
The following power law:
\begin{equation}
\mathcal{A}^{\ast}=\Phi^{0.7} \ ,
\label{Eq:MasterAPhi}
\end{equation}
provides an accurate forecast for the general trend of the contact area in the range $\Phi\in\left[10^{-2},10^0\right]$.
Note that only the contact area data for which $\mathcal{A}<\mathcal{A}_S$ is visible in Fig.~\ref{Fig:APhiDL}, with logarithmic scales, since for a SFP contact situation $\mathcal{A}=\mathcal{A}_S$ and the reduced area is $\mathcal{A}^{\ast}=0$.

\section{Conclusions}

\begin{figure}
\centering
\includegraphics[width=0.4\textwidth]{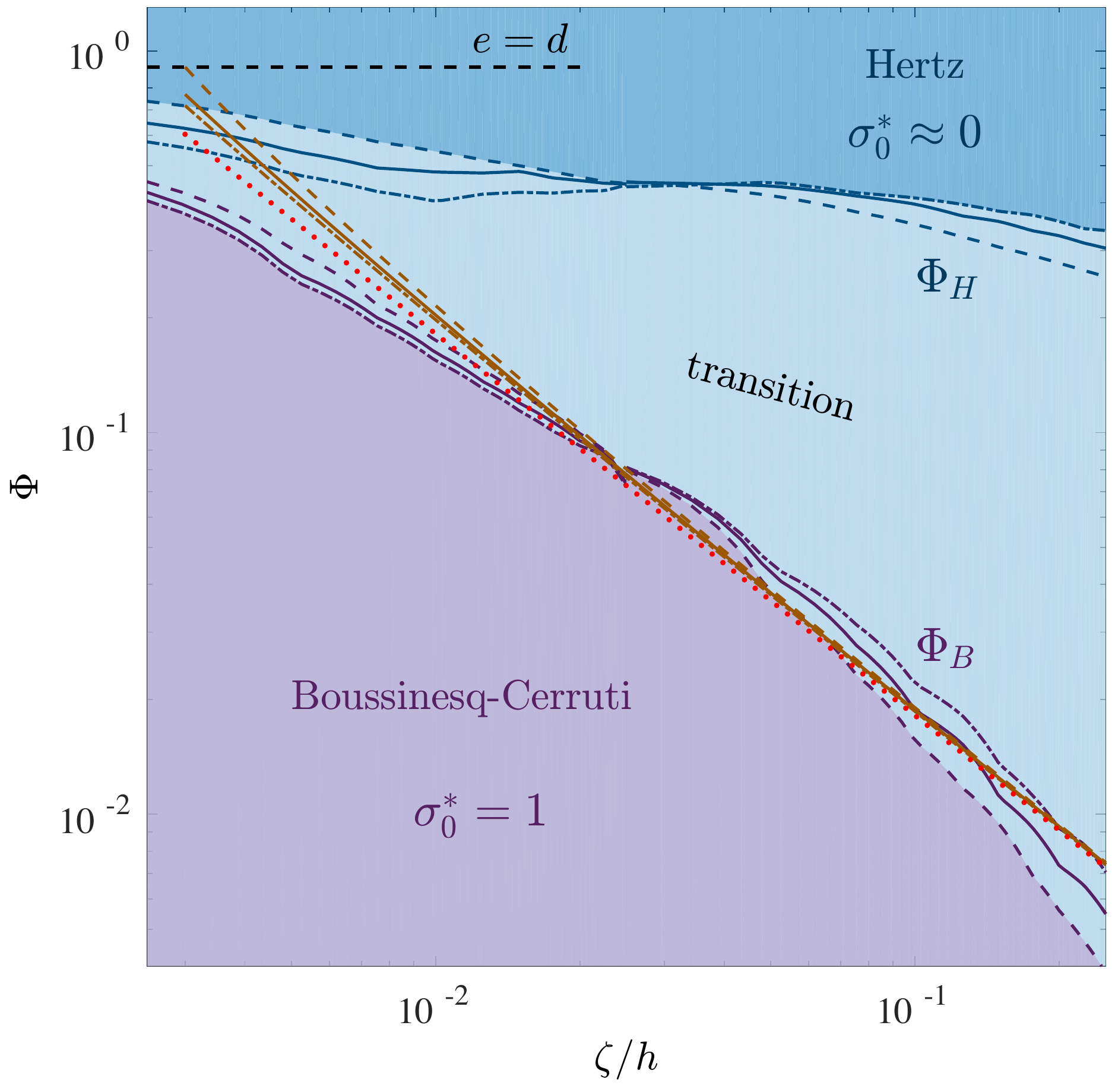}
\caption{Phase diagram of the elastic contact regime. The phase space is given in terms of $\Phi$ and the reduced indentation $\zeta/h$, for all the parameters given in Table~(\ref{Tab:Param}). The Hertz and soft-flat-punch (SFP) contact regions are bounded by the thresholds (\textbf{\textcolor{Violet}{violet}} and \textbf{\textcolor{Cyan}{blue}}, respectively), which were obtained numerically. The thresholds are depicted with different line styles for the different values of the Young's modulus ratio: [\textbf{- $\cdot$ -}] $E_p/E_l=10^{-3}$, [\textbf{---}] $E_p/E_l=1$ and [\textbf{- -}] $E_p/E_l=10^6$. For the SFP threshold, the approximation given by eqs.(\ref{Eq:rSzeta}-\ref{Eq:PhiSrS}) is shown as \textbf{\textcolor{brown}{brown}} curves for the 3 aforementioned Young's modulus ratii, whereas the large indentation approximation $\Phi_S= \pi d^2/\left(4\sqrt{3}R \zeta\right)$ corresponds to the overlapping [\textbf{\textcolor{red}{\large{$\cdots$}}}] dotted straight lines.}
\label{Fig:Phi0DLInd}
\end{figure}

In the present work, the elastic contact between a spherical lens and a patterned substrate (hexagonal lattice of pillars) has been studied in depth, using a superposition method of discrete pressure elements.
For relatively small indentations, a transition from a Hertzian elastic contact behavior towards a soft-flat-punch contact, considering a uniform displacement of the pillars, was observed when the surface fraction of pillars on the substrate is increased.
With the use of reduced variables for the stress at the center of the central pillar and the surface fraction occupied by the pillars, all the results collapse into a master curve, for which a rational function describes the passage between the aforementioned limiting contact regimes.
Additionally, a simple model to describe the behavior of a soft-flat-punch contact between the lens and a single elastic pillar has been presented as a function of the pillar geometry and the elastic properties of the lens and the pillar-substrate.
Special attention has been made on the effect of the ratio between the Young's moduli of both bodies.

Finally, from the reduced surface traction, two threshold values are obtained: the Hertzian $\Phi_H$ and the soft-flat-punch $\Phi_S$ limits.
To summarize these results, a phase diagram is depicted in Fig.\ref{Fig:Phi0DLInd}, which axes correspond to the surface fraction $\Phi$ and the reduced indentation $\zeta/h$.
The soft-flat-punch and Hertzian contact regions, which had been discerned from the numerical results, are shown.
In brief, for a given indentation $\zeta$, if the surface fraction $\Phi$ is greater than $\Phi_H$, the Hertzian contact offers a good approximation of the discrete contact phenomenon, whereas, if $\Phi$ is smaller than $\Phi_S$, the system must have only one pillar in contact with the lens.
In addition, also illustrated in Fig.\ref{Fig:Phi0DLInd}, an analytical expression for the soft-flat-punch threshold has been deducted from purely geometric considerations, which shows a good agreement with the numerical results, mainly in the large indentations regime.
In this region, a simple theoretical approximation is also represented, for which the effect of the elastic parameters seems to be negligible.
According to the theory developed by Greenwood~\cite{Greenwood1967,Greenwood1984} based on a statistical approach, the dimensionless parameter $\alpha=\mathcal{R}_a/\zeta$, which is the ratio between the $r.m.s.$ roughness of the surfaces in contact $\mathcal{R}_a$ and the total indentation $\zeta$, indicates the influence of the roughness on the contact mechanics.
For the hexagonal array of cylinders, since the sphere is considered to be smooth, $\mathcal{R}_a=h/2$, yielding the expression for the Greendwood dimensionless parameter:
\begin{equation}
\alpha=\dfrac{h}{2\zeta} \ .
\end{equation}
Considering a threshold value of $\alpha$, which has been given the empirical value of $\alpha=0.05$~\cite{Greenwood1984}, below which the Hertz theory can be applied with confidence, a vertical line in the phase diagram shown in Figure~\ref{Fig:Phi0DLInd} at the value of $\zeta/h=\left(2\alpha\right)^{-1}$ is found.
Even though this assumption should be valid for large indentations $\zeta/h\geq 10$, also corresponding to large contact regions, this procedure wipes out completely the impact of the surface fraction $\Phi$, at the indentation regime presented here.
Fortunately, a second dimensionless parameter $\mu$ is also proposed by Greenwood~\cite{Greenwood1967,Greenwood1984}, required to forecast precisely the contact behavior.
For rough nominally flat surfaces, $\mu$ depends only on the geometry, \textit{i.e.} the roughness, the asperity density and the $r.m.s.$ curvature of the asperities, and only generates a secondary effect on the stress and displacement fields when the contact zone spans infinitely compared to the nominal size of the asperities.
For the case presented herein, we expect that $\mu$ will take the form $\mu\sim\Phi\sqrt{R\mathcal{R}_a}/d$, which leads to an estimation of the second Greenwood dimensionless parameter:
\begin{equation}
\mu\sim\dfrac{\Phi}{d}\sqrt{\dfrac{Rh}{2}} \ ,
\end{equation}
where the effect of the surface fraction $\Phi$ appears directly.
\\
Thus, this analysis, which is a result of the qualitative comparison with the Greenwood's theory, indicates that $\Phi$ and $\zeta/h$ are the correct dimensionless parameters required to describe the passage from Hertzian to soft-flat-punch contact on periodically rough surfaces.

The methodology that we have developed allowed us to gain some insight on the discrete contact phenomenon.
Nevertheless, it only represents some first steps and several important issues remain to be addressed.
For instance, the case of large indentations, which are comparable in size with the height of the pillars $\zeta\sim h$, must be analyzed separately.
Under these circumstances, an intimate contact between the lens and the patterned surface, which has been studied with experiments \cite{Poulard2013,Verneuil2007,Jagota2010,Poulard2015}, may be triggered, and the role of adhesion should be taken into account.
Our efforts are currently being focused on this problem, and the corresponding analysis will be presented in a future publication.

\section{Acknowledgements and funding}

We wish to thank T. Salez for fruitful and very interesting discussions.
We have no competing interests.
This project was carried out thanks to funding by the ANR (ANR-11-BS04-0030 project).

\FloatBarrier

\footnotesize{
\bibliography{DiscreteContact} 

\begin{thebibliography}{10}

\bibitem{Johnson}
Johnson KL.
\newblock Contact mechanics.
\newblock Cambridge University Press; 1985.

\bibitem{restagno2002adhesion}
Restagno F, Crassous J, Cottin-Bizonne C, Charlaix E.
\newblock Adhesion between weakly rough beads.
\newblock Phys Rev E. 2002;65(4):042301.

\bibitem{Poulard2013}
Degrandi-Contraires E, Beaumont A, Restagno F, Weil R, Poulard C, Leger L.
\newblock Cassie-Wenzel-like transition in patterned soft elastomer adhesive
  contacts.
\newblock Eur Phys Lett. 2013;101(14001):1--6.

\bibitem{Greenwood1966}
Greenwood JA, Williamson JBP.
\newblock Contact of nominally flat surfaces.
\newblock Proc R Soc Lond A. 1966;295:300--319.

\bibitem{Greenwood1967}
Greenwood JA, Tripp JH.
\newblock The elastic contact of rough spheres.
\newblock J Appl Mech. 1967;89:153--159.

\bibitem{Greenwood1984}
Greenwood JA, Johnson KL, Matsubara E.
\newblock Surface roughness parameter in Hertz contact.
\newblock Wear. 1984;100:47--57.

\bibitem{Talke2010}
Li L, Etsion I, Talke FE.
\newblock Elastic?Plastic Spherical Contact Modeling Including Roughness
  Effects.
\newblock Tribol Lett. 2010;40:357--363.

\bibitem{Khonsari2014}
Beheshti A, Khonsari MM.
\newblock On the Contact of Curved Rough Surfaces: Contact Behavior and
  Predictive Formulas.
\newblock J Appl Mech. 2014;81(111004):1--15.

\bibitem{Bush1975}
Bush AW, Gibson RD, Thomas TR.
\newblock The elastic contact of a rough surface.
\newblock Wear. 1975;35:87--111.

\bibitem{Persson2006}
Persson BNJ.
\newblock Contact mechanics for randomly rough surfaces.
\newblock Surf Sci Rep. 2006;61:201--227.

\bibitem{Johnson1984}
Johnson KL, Greenwood JA, Higginson JG.
\newblock The contact of elastic regular wavy surfaces.
\newblock Int J Mech Sci. 1984;27(6):383--396.

\bibitem{Block2008}
Block JM, Keer LM.
\newblock Periodic contact problems in plane elasticity.
\newblock J Mech Mater Struct. 2008;3(7):1207--1237.

\bibitem{Archard1953}
Archard JF.
\newblock Contact and Rubbing of Flat Surfaces.
\newblock J Appl Phys. 1953;24(981):1--8.

\bibitem{Murarash2011}
Murarash B, Itovich Y, Varenberg M.
\newblock Tuning elastomer friction by hexagonal surface patterning.
\newblock Soft Matter. 2011;7(12):5553--5557.

\bibitem{Brormann2013}
Brormann K, Barel I, Urbakh M, R B.
\newblock Friction on a Microstructured Elastomer Surface.
\newblock Tribol Lett. 2013;50(1):3--15.

\bibitem{Kligerman2014}
Kligerman Y, Varenberg M.
\newblock Elimination of Stick-Slip Motion in Sliding of Split or Rough
  Surface.
\newblock Tribol Lett. 2014;53(2):395--399.

\bibitem{Tsipenyuk2014}
Tsipenyuk A, Varenberg M.
\newblock Use of biomimetic hexagonal surface texture in friction against
  lubricated skin.
\newblock J R Soc Interface. 2014;11:1--13.

\bibitem{Romero2014}
Romero V, Wandersman E, Debr\'{e}geas G, Prevost A.
\newblock Probing Locally the Onset of Slippage at a Model Multicontact
  Interface.
\newblock Phys Rev Lett. 2014;112(094301):1--5.

\bibitem{Mittal}
Mittal KL.
\newblock Polymer surface modification: Relevance to adhesion, Volume 3.
\newblock CRC Press; 2004.

\bibitem{Kozlova2006}
Kozlova N, Santore MM.
\newblock Manipulation of Micrometer-Scale Adhesion by Tuning Nanometer-Scale
  Surface Features.
\newblock Langmuir. 2006;22(3):1135--1142.

\bibitem{Poulard2011}
Poulard C, Restagno F, Weil R, Leger L.
\newblock Mechanical tuning of adhesion through micro-patterning of elastic
  surfaces.
\newblock Soft Matter. 2011;101(7):2543--2551.

\bibitem{Tamelier2012}
Tamelier J, Chary S, Turner KL.
\newblock Vertical Anisotropic Microfibers for a Gecko-Inspired Adhesive.
\newblock Langmuir. 2012;28(23):8746--8752.

\bibitem{gorb2007}
Gorb S, Varenberg M, Peressadko A, Tuma J.
\newblock Biomimetic mushroom-shaped fibrillar adhesive microstructure.
\newblock Journal of The Royal Society Interface. 2007;4(13):271--275.

\bibitem{bhushan2011}
Bhushan B, Jung YC.
\newblock Natural and biomimetic artificial surfaces for superhydrophobicity,
  self-cleaning, low adhesion, and drag reduction.
\newblock Progress in Materials Science. 2011;56(1):1--108.

\bibitem{hui2004}
Hui CY, Glassmaker N, Tang T, Jagota A.
\newblock Design of biomimetic fibrillar interfaces: 2. Mechanics of enhanced
  adhesion.
\newblock Journal of The Royal Society Interface. 2004;1(1):35--48.

\bibitem{zhou2013}
Zhou M, Pesika N, Zeng H, Tian Y, Israelachvili J.
\newblock Recent advances in gecko adhesion and friction mechanisms and
  development of gecko-inspired dry adhesive surfaces.
\newblock Friction. 2013;1(2):114--129.

\bibitem{crosby2005}
Crosby AJ, Hageman M, Duncan A.
\newblock Controlling polymer adhesion with ``pancakes''.
\newblock Langmuir. 2005;21(25):11738--11743.

\bibitem{murphy2009}
Murphy MP, Kim S, Sitti M.
\newblock Enhanced adhesion by gecko-inspired hierarchical fibrillar adhesives.
\newblock ACS applied materials \& interfaces. 2009;1(4):849--855.

\bibitem{benz2006deformation}
Benz M, Rosenberg KJ, Kramer EJ, Israelachvili JN.
\newblock The deformation and adhesion of randomly rough and patterned
  surfaces.
\newblock The Journal of Physical Chemistry B. 2006;110(24):11884--11893.

\bibitem{greiner2007adhesion}
Greiner C, del Campo A, Arzt E.
\newblock Adhesion of bioinspired micropatterned surfaces: effects of pillar
  radius, aspect ratio, and preload.
\newblock Langmuir. 2007;23(7):3495--3502.

\bibitem{Verneuil2007}
Verneuil E, Ladoux B, Buguin A, Silberzan P.
\newblock Adhesion on microstructured surfaces.
\newblock J Adhesion. 2007;83:449--472.

\bibitem{Hisler2013}
Hisler V, Palmieri M, Le~Houerou V, Gauthier C, Nardin M, Vallat MF, et~al.
\newblock Scale invariance of the contact mechanics of micropatterned elastic
  substrates.
\newblock Int J Adhesion Adhesives. 2013;45:144--149.

\bibitem{Jagota2010}
Nadermann N, Ning J, Jagota A, Hui CY.
\newblock Active switching of adhesion in a film-terminated fibrillar
  structure.
\newblock Langmuir. 2010;26(19):15464--15471.

\bibitem{Love}
Love AEH.
\newblock A treatise on the mathematical theory of elasticity.
\newblock 4th ed. Cambridge University Press; 1952.

\bibitem{Sneddon1945}
Harding JW, Sneddon IN.
\newblock The elastic stresses produced by the indentation of the plane surface
  of a semi-infinite elastic solid by a rigid punch.
\newblock Math Proc Cambridge. 1945;41(1):16--26.

\bibitem{Maugis}
Maugis D.
\newblock Contact, adhesion and rupture of elastic solids.
\newblock Springer; 1999.

\bibitem{Gecit1986}
Gecit MR.
\newblock Axisymmetric contact problem for a semi-infinite cylinder and a half
  space.
\newblock Int J Engng Sci. 1986;24(8):1245--1256.

\bibitem{SM}
See Supplemental Material at [URL will be inserted by publisher] for details on
  the numerical method;.

\bibitem{Guidoni2010}
Guidoni GM, Schillo D, Hangen U, Castellanos G, Arzt E, McMeeking RM, et~al.
\newblock Discrete contact mechanics of a fibrillar surface with backing layer
  interactions.
\newblock J Mech Phys Solids. 2010;58:1571--1581.

\bibitem{Jagota2007}
Noderer WL, Shen L, Vajpayee S, Glassmaker NJ, Jagota A, Hui CY.
\newblock Enhanced adhesion and compliance of film-terminated fibrillar
  surfaces.
\newblock Proc R Soc A. 2007;463:2631--2654.

\bibitem{Li1997}
Li J, Chou TW.
\newblock Elastic field of a thin-film/substrate system under an axisymmetric
  loading.
\newblock Int J Solids Struct. 1997;34(35--36):4463--4478.

\bibitem{Poulard2015}
Dies L, Restagno F, Weil R, Leger L, Poulard C.
\newblock Role of adhesion between asperities in the formation of elastic
  solid/solid contacts.
\newblock Eur Phys J E. 2015;38(130):1--8.

\end{thebibliography}


\begin{thebibliography}{1}

\bibitem{Love1929}
Love AEH.
\newblock The stress produced in a semi-infinite solid by pressure on part of
  the boundary.
\newblock Phil Trans R Soc Lond A. 1929;667:377--420.

\bibitem{Johnson}
Johnson KL.
\newblock Contact mechanics.
\newblock Cambridge University Press; 1985.

\end{thebibliography}
\bibliographystyle{vancouver} 
}

\end{document}


\title{{\normalsize Supplementary material of the paper:}\\ Stress concentration in periodically rough hertzian contact: Hertz to soft-flat-punch transition}
\author[1,2,$^{\dagger}$]{Ren\'{e} Ledesma-Alonso}
\author[1]{Elie Rapha\"{e}l}
\author[2]{Liliane L\'{e}ger}
\author[2]{Frederic Restagno}
\author[2]{Christophe Poulard}
\affil[1]{Laboratoire de Physico-Chimie Th\'{e}orique, UMR CNRS 7083 Gulliver, ESPCI ParisTech, PSL Research University, 10 Rue Vauquelin, 75005 Paris, France}
\affil[2]{Laboratoire de Physique des Solides, CNRS, Univ. Paris-Sud, Universit\'e Paris-Saclay, 91400 Orsay, France}\affil[$^{\dagger}$]{e-mail: rene.ledesma-alonso@espci.fr}
\date{\today}
\maketitle

\section{Pillar array and symmetry}
\label{App:Pillararray}

\begin{figure}
\centering
\subfigure[ ]{
\centering
\includegraphics[width=0.4\textwidth]{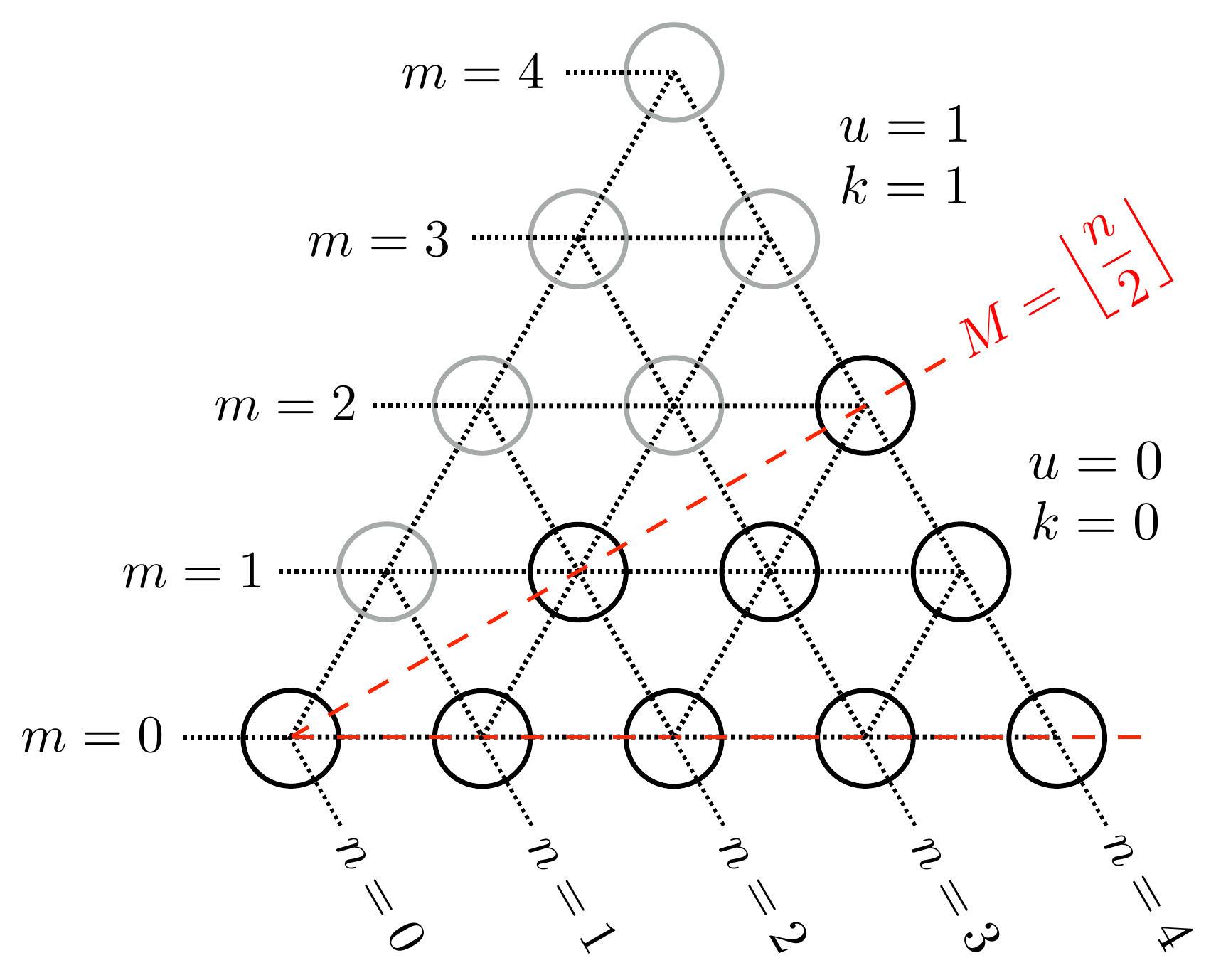}
\label{Fig:Indices}
}
\hspace{5mm}
\subfigure[]{
\centering
\includegraphics[width=0.45\textwidth]{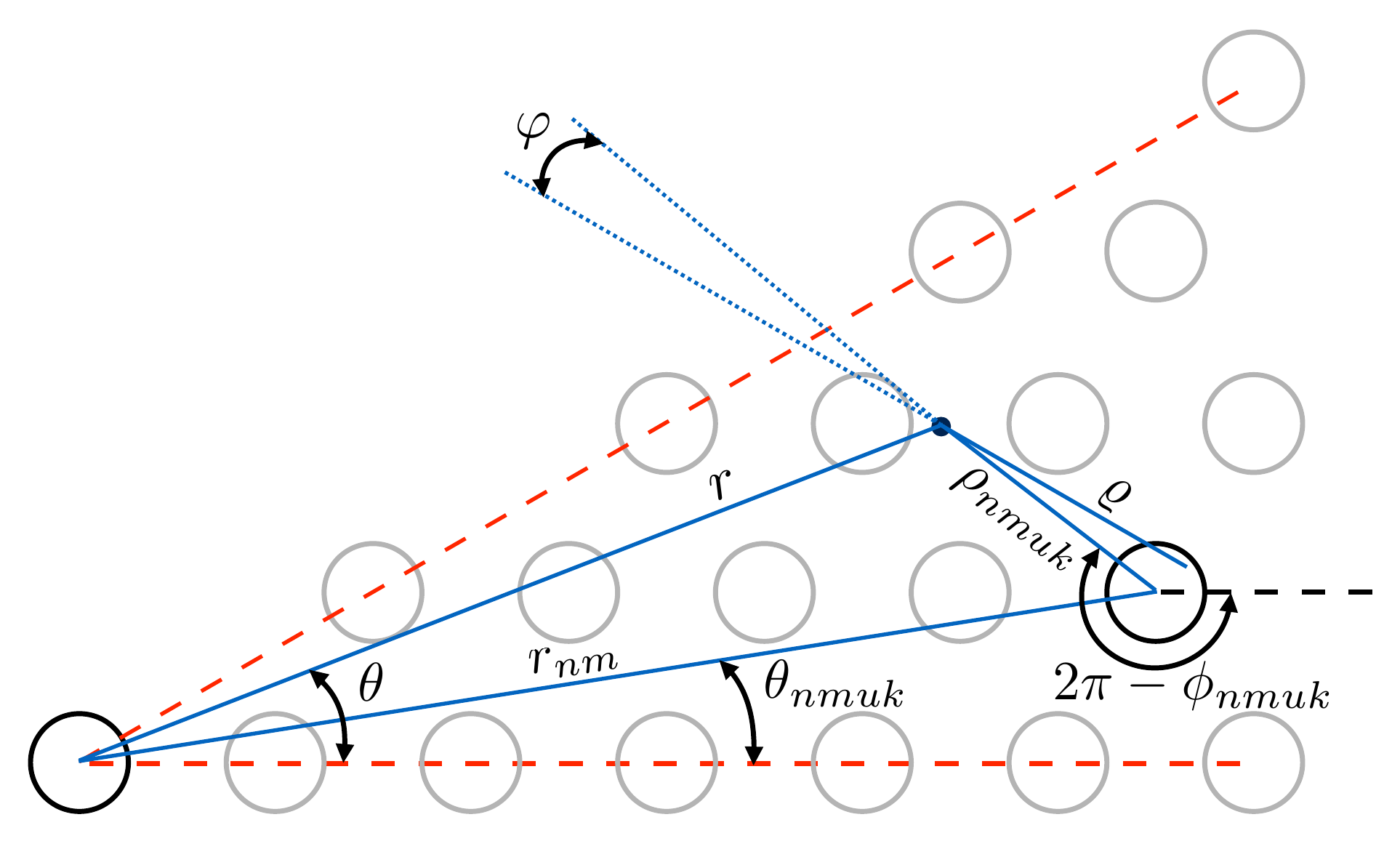}
\label{Fig:Relative}
}
\caption{\subref{Fig:Indices} Pillar indexation and \subref{Fig:Relative} relative position of a point $\left(r,\theta\right)$ with respect to an arbitrary position at the top surface of the $nmuk$-pillar. In both figures, the primitive cell is delimited by the [\textcolor{red}{$- -$}] discontinuous lines.}
\end{figure}

At the substrate, each cylindrical pillar is fixed and placed at the intersection of two lines, the first described by the function $y=-\sqrt{3}\left[x-n \, e\right]$ and the second by $y=\left[\sqrt{3}/2\right]m \, e$, where $n$ and $m$ are integers that enumerate each of the lines, as shown in Fig.~\ref{Fig:Indices}, and $e$ is the pitch distance between pillars.
Thus, the center of the $nm$-pillar is placed at the coordinates $x_{nm} =\left[2n-m\right]e/2$ and $y_{nm}=\sqrt{3}\, m \, e/2$.
In addition, since the spherical lens is centered above the center of the $n=m=0$ pillar and due to the symmetry of a hexagonal pillar lattice, which has 6 rotation symmetries and 6 reflection symmetries, the primitive cell of our system is a circular sector with central angle $\pi/6$.
The primitive cell is also shown in Fig.~\ref{Fig:Indices}.
As a consequence, the ranges of the indices $n$ and $m$ are coupled and restricted to $n\in \left[0,N\right]$ and $m\in\left[0,M\right]$, and we must also take into account two new indices $u$ and $k$, which are also restricted to $u\in\left[0,5\right]$ and $k\in\left[0,K\right]$, and stand for the aforementioned symmetries of the system.
The value of $N$ is strictly related to the contact radius, whereas $M$ depends on the index $n$ and $K$ depends on both indices $n$ and $m$, which relationships are given in Table~\ref{Tab:Ranges}.

\begin{table}[htbp]
\begin{center}
    \vspace*{5mm}
    \begin{tabular}{|c|c|c|c|c|}
    \hline
    $n$ &$M\left(n\right)$ & $m$ & $K\left(n,m\right)$ \\ \hline
    $n<2$ & 0 & & 0 \\ \hline
    $n=2$ & 1 & & 0 \\ \hline
    $n>2$ \& odd & $\left(n-1\right)/2$ & $0$ & 0 \\ 
    & & $1\leq m\leq M$ & 1 \\ \hline
    $n>2$ \& even & $n/2$ & $0$ & 0 \\ 
    & & $1\leq m\leq M-1$ & 1 \\ 
    & & $M$ & 0 \\ \hline
    \end{tabular}
\end{center}
\caption{Ranges of the indices $n$ and $m$, and their relationship with the limit values $M$ and $K$, depending on $n$ and $m$.}
\label{Tab:Ranges}
\end{table}

In a polar system $\left(r,\theta\right)$, we can define the position of each pillar in terms of the four indices:
\begin{align}
r_{nm} &=\dfrac{e}{2} \sqrt{\left(2n-m\right)^2+3m^2} \ , \notag \\
\theta_{nmuk} &=\dfrac{\pi u}{3}+\left(-1\right)^k \theta_{nm} \ , \notag \\
\theta_{nm} &=\arctan\left(\dfrac{\sqrt{3}m}{2n-m}\right) \ ,
\label{Eq:Polpos}
\end{align}
where $r_{nm}$ and $\theta_{nmuk}$ are radial and angular coordinates, respectively.
According to the aforementioned description, and related to the indices $n=m=0$, a pillar with coordinates $r_{00}=0$ and $\theta_{00}=\theta_{00uk}=0$ is located at the origin of the coordinate system, which is also aligned with the lowest point of the spherical lens.

\section{Displacement and indentation}

According to the introduced indexation, the pillars/substrate combined surface is described by the following expression:
\begin{align}
g^{\circ}\left(r,\theta\right) &=h \left[1-H\left(\rho_{00}-\dfrac{d}{2}\right)\right]
+\sum_{n=1}^{N}\sum_{m=0}^{M}\sum_{u=0}^{5}\sum_{k=0}^{K}\left[1-H\left(\rho_{nmuk}-\dfrac{d}{2}\right)\right] \ , \notag
\end{align}
where $H$ is a Heaviside step function.
As it is depicted in Fig.~\ref{Fig:Relative}, $\rho_{nmuk}$ is the radial distance of the point $\left(r,\theta\right)$ relative to the center of the $nmuk$-pillar, which reads:
\begin{equation}
\rho_{nmuk}=\sqrt{r^2+r_{nm}^2-2r r_{nm} \cos\left(\theta-\theta_{nmuk}\right)} \ .
\label{Eq:Distnmuk}
\end{equation}

Once the load is applied and the system has attained equilibrium, the total displacement field $w$ is composed of:
\begin{equation}
w\left(r,\theta\right)=w_l\left(r,\theta\right)+w_s\left(r,\theta\right)+\dfrac{\delta_{nmuk}}{h} \, g^{\circ}\left(r,\theta\right) \ ,
\label{Eq:DispDecomp}
\end{equation}
where $w_l$ and $w_s$ are the normal displacement of the lens and substrate, and $\delta_{nmuk}$ is the change of height of the $nmuk$-pillar.

Due to the discrete nature of the contact, the normal displacement of each surface can be decomposed as follows:
\begin{align}
w_j\left(r,\theta\right) &=\dfrac{\gamma_j}{\pi}\left[ W_{00}+\sum_{n=1}^{N} \sum_{m=0}^{M} \sum_{u=0}^{5} \sum_{k=0}^{K} W_{nmuk}\right] \ , \notag \\
W_{nmuk} &=\int_{\varphi_1}^{\varphi_2}\int_{\varrho_1}^{\varrho_2} \sigma\left(r_{nm}\ ,\theta_{nmuk},\varrho,\varphi\right) d\varrho d\varphi \ ,
\label{Eq:DDisc}
\end{align}
where $j=l,s$ refers to the lens or substrate, $W_{nmuk}$ is the reduced displacement due to the contact with the $nmuk$-pillar, for the center pillar $W_{00}=W_{00uk}$ and $\left(\varrho,\varphi\right)$ are the relative coordinates of the center of the $nmuk$-pillar with respect to the point $\left(r,\theta\right)$, as shown in Fig.~\ref{Fig:Relative}.

Also owing to the symmetry of the system, the change of height becomes $\delta_{nm}=\delta_{nmuk}$, and since the compression of the $nmuk$-pillar is approximated by the Hooke's law, it is equal to:
\begin{equation}
\delta_{nm}=\dfrac{4h}{\pi d^2E_p}F_{nm} \ .
\label{Eq:DCyl}
\end{equation}
In turn, the lens and substrate indentations are given by the corresponding displacements at the origin $r=0$, thus $\zeta_{l,s}=w_{l,s}\left(0,\theta\right)$, and the total indentation $\zeta$ corresponds to:
\begin{equation}
\zeta=\zeta_l+\zeta_s+\delta_{00} \ .
\label{Eq:TotIdent}
\end{equation}

The total load is obtained by adding the contribution of each pillar as follows:
\begin{equation}
F_T=F_{00}+6\sum_{n=1}^{N} \sum_{m=0}^{M} \sum_{k=0}^{K} F_{nm0k} \ ,
\label{Eq:FDisc}
\end{equation}
where $F_{nm}=F_{nmuk}$ is the load exerted over the $nmuk$-pillar in the contact region, which reads:
\begin{equation}
F_{nm}=\int_{0}^{2\pi}\int_{0}^{d/2} \sigma\left(r_{nm},\theta_{nm},\rho,\phi\right)\rho d\rho d\phi \ ,
\label{Eq:FCyl}
\end{equation}
whereas $\rho=\rho_{nmuk}$ and $\phi=\phi_{nmuk}$ represent the radial and angular coordinates at the surface of the $nmuk$-pillar, relative to the pillar center $\left(r_{nm},\theta_{nmuk}\right)$ and restricted to the ranges $\rho\in\left[0,d/2\right]$ and $\phi\in\left[0,2\pi\right]$.
In order to transform $\sigma\left(r,\theta\right)$ into $\sigma\left(r_{nm},\theta_{nmuk},\rho,\phi\right)$, one should express $r$ and $\theta$ for a given relative position $\left(\rho,\phi\right)$ and a pillar $\left(r_{nm},\theta_{nmuk}\right)$ using the following expressions:
\begin{align}
r &=\sqrt{\rho^2+r_{nm}^2+2\rho r_{nm} \cos\left(\phi-\theta_{nmuk}\right)} \ , \notag \\
\theta &=\arcsin\left(\dfrac{\rho\sin\left(\pi-\phi\right)+r_{nm}\sin\left(\theta_{nmuk}\right)}{r}\right) \ ,
\label{Eq:CylRel}
\end{align}
which are deduced form Fig.\ref{Fig:Relative}.

Finally, at the local contact region of each pillar, we have $\Delta\left(r,\theta\right)=0$ and $G\left(r,\theta\right)=1$, and thus the total displacement field $w$ is related to the total indentation $\zeta$ as follows:
\begin{equation}
w\left(r,\theta\right)=\zeta-\left[R-\sqrt{R^2-r^2}\right] \ .
\label{Eq:TotDisp}
\end{equation}
By imposing a total indentation in eq.\eqref{Eq:TotDisp}, and gathering the displacement contributions given in eqs.~\eqref{Eq:DDisc} and \eqref{Eq:DCyl} to assemble eq.\eqref{Eq:DispDecomp}, the stress field $\sigma$ at the contact region, \textit{i.e.} the top of the pillars by taking $G\left(r,\theta\right)=1$, is recovered numerically using a superposition method of discrete pressure elements, which is detailed in section~\ref{App:DiscretePressure}.

%
%

\section{Discrete pressure elements}
\label{App:DiscretePressure}

\begin{figure}
\centering
\includegraphics[width=0.4\textwidth]{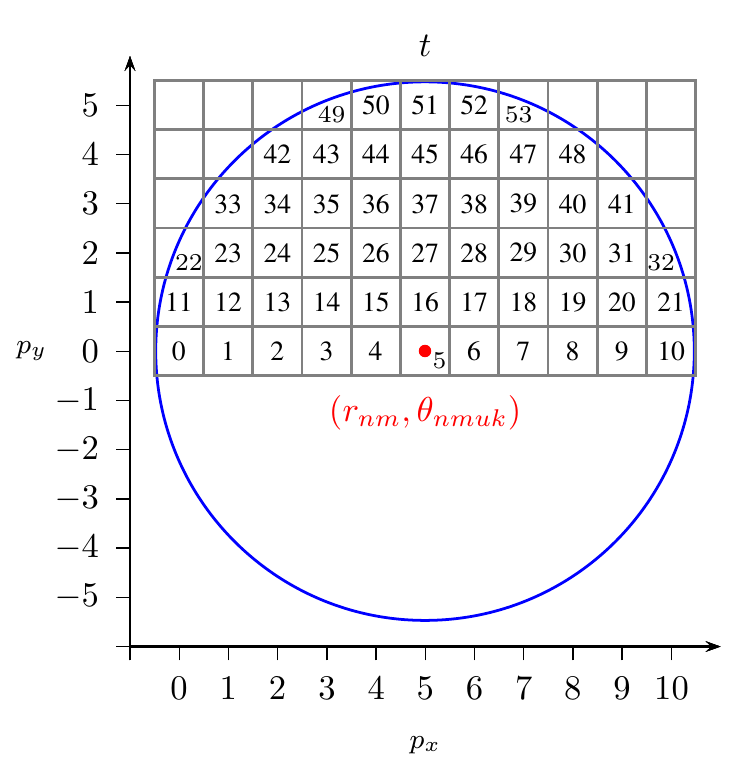}
\caption{Discretization of a pillar into $T=53$ square elements, with $P=11$ elements along the $p_y=0$ axis.}
\label{Fig:Discrete}
\end{figure}


We will now introduce the variable $q$, defined as:
\begin{equation}
q=-1+\sum_{n=0}^{N} \sum_{n=0}^{M} \left\{1-H\left(r_{nm}-\left[a+\dfrac{d}{2}\right]\right)\right\} \ ,
\end{equation}
where $H$ is a Heaviside step function, and the new variable presents the range $q\in\left[0,Q\right]$.
This variable allows us to count the total number of pillars $Q$ within the contact region, and to transform the double finite series into a single series:
\begin{align}
\sum_{n=0}^{N} \sum_{n=0}^{M} & \rightarrow \sum_{q=0}^{Q} & \text{ and } & & \sum_{n=1}^{N} \sum_{n=0}^{M} & \rightarrow \sum_{q=1}^{Q} \ ,
\end{align}
and to compact the two indices notation $nm$ into a single index $q$, for the variables introduced in the previous section, which now read: $r_q$ and $\theta_{quk}$ radial and angular position, $F_{quk}$ force, $\delta_{quk}$ change of height, $W_{quk}$ and reduced displacement of the $quk$-pillar.

Now, consider a square region of area $d^2$, centered at the position of a pillar and divided into $P=d/b$ elements per side, which generates square subregions of area $b^2$.
Introducing the variable $t$, defined as:
\begin{align}
t &=-1+\sum_{j=0}^{p_y} \sum_{i=0}^{p_x} \left[1-H\left(r_{ij}^{\ast}-\dfrac{d}{2}\right)\right] \ , \notag \\
r_{ij}^{\ast} &=b\sqrt{\left(i-\dfrac{P-1}{2}\right)^2+j^2} \ .
\end{align}
where $p_x$ and $p_y$ are integers in the ranges $p_x\in\left[0,P-1\right]$ and $p_y\in\left[0,\left(P-1\right)/2\right]$, and $r_{ij}^{\ast}$ is the distance between the element $\left(p_x,p_y\right)$ and the center of the pillar $\left(r_q,\theta_{quk}\right)$.
Considering this discretization, we count a total of $T$ subregions within the half-pillar top surface, as depicted in Fig.\ref{Fig:Discrete}.
The center of each square $t$ is placed at the coordinates $\left(x_t,y_t\right)$, relative to the center of the $quk$-pillar, which read:
\begin{align}
x_t&=b\left[-\dfrac{P+1}{2}+\sum_{i=0}^{p_x} 1\right]
\ , & y_t&=b\, p_y \ .
\end{align}
The discrete axis $p_x$ and $p_y$ are parallel and perpendicular to the vector pointing from the origin of the system $r=0$ towards the pillar centered at $\left(r_q,\theta_{quk}\right)$.


In addition, the symmetry of the system configuration allows us to identify that the stress at the $quk$-pillar is independent of the angular position. 
Therefore, we should only compute the stress at each $q$-pillar, centered at the position $\left(r_q,\theta_q\right)$, with $\theta_q=\theta_{q00}$.
Each square $t$ is subjected to a uniform pressure $\sigma_{qt}$, which generates a contribution to the reduced displacement $W_{quk}$, already defined in eq.\eqref{Eq:DDisc}.
Therefore, at a general point $\left(r,\theta\right)$, the displacement generated by the contact of the lens with the $quk$-pillar is:
\begin{align}
W_{quk} &=\sum_{t=0}^{P-1} \sigma_{qt} D_{qukt0}+\sum_{t=P}^{T} \sigma_{qt} \left(\sum_{i=0}^{1} D_{qukti}\right) \ ,
\end{align}
where the function $D$ is defined~\cite{Love1929,Johnson} as:
\begin{align}
D_{qukti} &=\sum_{j=1}^2\left\{
\left[X+\left(-1\right)^j \dfrac{b}{2}\right]\ln\left(\dfrac{\left[Y+\left(-1\right)^j \dfrac{b}{2}\right]+\sqrt{\left[Y+\left(-1\right)^j \dfrac{b}{2}\right]^2+\left[X+\left(-1\right)^j \dfrac{b}{2}\right]^2}}{\left[Y-\left(-1\right)^j \dfrac{b}{2}\right]+\sqrt{\left[Y-\left(-1\right)^j \dfrac{b}{2}\right]^2+\left[X+\left(-1\right)^j \dfrac{b}{2}\right]^2}}\right)\right. \notag \\
& \qquad \qquad \qquad \left.
+\left[Y+\left(-1\right)^j \dfrac{b}{2}\right]\ln\left(\dfrac{\left[X+\left(-1\right)^j \dfrac{b}{2}\right]+\sqrt{\left[Y+\left(-1\right)^j \dfrac{b}{2}\right]^2+\left[X+\left(-1\right)^j \dfrac{b}{2}\right]^2}}{\left[X-\left(-1\right)^j \dfrac{b}{2}\right]+\sqrt{\left[Y+\left(-1\right)^j \dfrac{b}{2}\right]^2+\left[X-\left(-1\right)^j \dfrac{b}{2}\right]^2}}\right)\right\},
\label{Eq:Dqukt}
\end{align}
with:
\begin{align}
X &= \bigg\vert r_q+x_t-r\cos\left(\theta_{quk}-\theta\right)\bigg\vert \ , &
Y &= \bigg\vert \left(-1\right)^i y_t+r\sin\left(\theta_{quk}-\theta\right)\bigg\vert \ .
\label{Eq:XY}
\end{align}
At the contact between the lens and the $quk$-pillar, the exerted mutual load is calculated from the sum:
\begin{equation}
F\left(r_q,\theta_{quk}\right)=b^2\left[\sum_{t=0}^{P-1} \sigma_{qt}+2\sum_{t=P}^{T} \sigma_{qt} \right] \ .
\end{equation}

From eq.\eqref{Eq:DispDecomp}, the addition of the pillar height change and the displacements of the lens and the substrate gives the total displacement, which reads:
\begin{align}
w\left(r,\theta\right) &=\dfrac{4 h}{\pi E_p}\left(\dfrac{b}{d}\right)^2 \left\{
\vphantom{
\sum_{q=1}^{Q} \sum_{u=0}^{5} \sum_{k=0}^{K} \left[\sum_{t=0}^{P-1}
\sum_{t=P}^{T} \sigma_{qt} \left(\sum_{i=0}^{1} D_{qukti}\right)\right]}
\left[1-H\left(r-\dfrac{d}{2}\right)\right] \left[\sum_{t=0}^{P-1} \sigma_{qt}+2\sum_{t=P}^{T} \sigma_{qt} \right]\right. \notag \\
& \qquad \quad \qquad \qquad \quad \left.+\sum_{q=1}^{Q} \sum_{u=0}^{5} \sum_{k=0}^{K} \left[1-H\left(\rho_{quk}-\dfrac{d}{2}\right)\right] \left[\sum_{t=0}^{P-1} \sigma_{qt}+2\sum_{t=P}^{T} \sigma_{qt} \right]\right\} \notag \\
& \qquad +\left(\dfrac{\gamma_l+\gamma_s}{\pi}\right)\left\{\sum_{t=1}^{T} \sigma_{0t} D_{000t0}
+\sum_{q=1}^{Q} \sum_{u=0}^{5} \sum_{k=0}^{K} \left[\sum_{t=0}^{P-1} \sigma_{qt} D_{qukt0}+\sum_{t=P}^{T} \sigma_{qt} \left(\sum_{i=0}^{1} D_{qukti}\right)\right]\right\} \ ,
\label{Eq:SigSyst}
\end{align}
recalling that $r=\rho_{00uk}$ and $\theta=\phi_{00uk}$.

Therefore, to calculate $\sigma_{qt}$, we must furnish the values of $w$ at the contact region, \textit{i.e.} the discrete positions $\left(x_t,y_t\right)$ at each pillar.
Firstly, since the same number of locations $qt$ is required, we substitute the radial and angular coordinates:
\begin{align}
r &=\sqrt{\left(r_q+x_t\right)^2+y_t^2} \ , \notag \\
\theta &=\theta_{quk}+\arctan\left(\dfrac{y_t}{r_q+x_t}\right) \ ,
\label{Eq:CompCoord}
\end{align}
within eq.\eqref{Eq:TotDisp} and with $u=k=0$ to acquire $w\left(r,\theta\right)$, and then, within eqs.(\ref{Eq:Dqukt}-\ref{Eq:XY}) to compute $D_{qukti}$, for the complete ranges $q\in\left[0,Q\right]$, $u\in\left[0,5\right]$, $k\in\left[0,K\right]$ and $t\in\left[0,T\right]$.
Afterwards, we include both terms in the right-hand and left-hand sides of eq.\eqref{Eq:SigSyst}, together with the coordinates $(r,\theta)$ from eq.\eqref{Eq:CompCoord}, and proceed to recover $\sigma_{qt}$ by solving numerically the resulting system of linear equations.

\section{Contact region estimate}
\label{App:ContactRegion}

An iterative bisection-like method has been implemented to determine the contact radius $a$.
Its implementation consists in performing the following steps:

\begin{enumerate}
\item As initial condition for $i=0$, use the overlap radius at which $f^{\circ}\left(r\right)=h-\zeta$, as a guess of the contact radius $a_i$.
\item Compute the number of pillars $Q_i$ in contact by adding one by one, each pillar which accomplishes the condition $r_{nm}/\left(a_i+d/2\right)<1$.
\item Find the displacement field $w$ within the proposed contact region, using eq.\eqref{Eq:TotDisp}.
\item Determine the stress field $\sigma$ by applying the method described in App.\ref{App:DiscretePressure}.
\item If there are zones showing negative stress values $\sigma<0$, identify the closest position $a_{i+1}$ (with respect to $r=0$) at which $\sigma\leq 0$.
\item Repeat the process from step 2 until $\vert a_{i+1}-a_i\vert<b$, which indicates that the contact radius had converged to a constant value, within the resolution of the numerical method.
\end{enumerate}

\FloatBarrier

\footnotesize{
\bibliography{DiscreteContact} 
\bibliographystyle{vancouver} 
}